\documentclass[twocolumn,aps,prd,amsmath,amssymb,preprintnumbers,longbibliography]{revtex4-1}
\usepackage{amsmath} \usepackage{amsfonts} \usepackage{amssymb}
\usepackage{bbm}
\usepackage{epsfig}
\usepackage{graphics}
\usepackage{graphicx}
\usepackage{titlesec}
\usepackage{mathtools}
\usepackage{environ}
\usepackage{dsfont}
\usepackage{version}

\usepackage{natbib}
\pdfoutput=1
\usepackage[colorlinks=true]{hyperref}
\hypersetup{urlcolor=black,linkcolor=black,citecolor=black}
\usepackage[toc,page]{appendix}
\usepackage{enumitem}
\makeatletter
\renewcommand*\env@matrix[1][c]{\hskip -\arraycolsep
  \let\@ifnextchar\new@ifnextchar
  \array{*\c@MaxMatrixCols #1}}
\makeatother

\newcommand{\be}{\begin{equation}}
\newcommand{\ee}{\end{equation}}
\newcommand{\ba}{\begin{align}}
\newcommand{\ea}{\end{eqnarray}}
\newcommand{\nn}{\nonumber}

\newcommand{\2}{^{2}} 

\titleformat{\subsection}[block]{\normalfont\bfseries}{\thesubsection.}{1ex}{}
\titlespacing{\subsection}{0pt}{10pt}{1pt}[0pt]
\titleformat*{\section}{\large\bfseries}
\renewcommand{\thesubsection}{\arabic{subsection}}

\usepackage{todonotes}
\usepackage{braket}

\catcode`\@=11
\def\lsim{\mathrel{\mathpalette\@versim<}}
\def\gsim{\mathrel{\mathpalette\@versim>}}
\def\@versim#1#2{\vcenter{\offinterlineskip
\ialign{$\m@th#1\hfil##\hfil$\crcr#2\crcr\sim\crcr } }}
\catcode`\@=12

\newcommand{\p}{\partial}

\newcommand{\A}{\mathcal{A}}

\newcommand{\al}[1]{\begin{align}#1\end{align}}
\newcommand{\bp}{\begin{pmatrix}}
\newcommand{\ep}{\end{pmatrix}}

\newcommand{\df}{\text{d}}

\newcommand{\bs}[1]{\boldsymbol}

\graphicspath{{./figs/}}

\definecolor{refkey}{rgb}{0,0,1}
\definecolor{labelkey}{rgb}{0,1,0}

\begin{document}


\title{\LARGE Pregeometry and euclidean quantum gravity}



\author{Christof Wetterich}

\affiliation{Institut  f\"ur Theoretische Physik\\
Universit\"at Heidelberg\\
Philosophenweg 16, D-69120 Heidelberg}



\begin{abstract}
Einstein's general relativity can emerge from pregeometry, with the metric composed of more fundamental fields. We formulate euclidean pregeometry as a $SO(4)$ - Yang-Mills theory. In addition to the gauge fields we include a vector field in the vector representation of the gauge group. The gauge - and diffeomorphism - invariant kinetic terms for these fields permit a well-defined euclidean functional integral, in contrast to metric gravity with the Einstein-Hilbert action. The propagators of all fields are well behaved at short distances, without tachyonic or ghost modes. The long distance behavior is governed by the composite metric and corresponds to general relativity. In particular, the graviton propagator is free of ghost or tachyonic poles despite the presence of higher order terms in a momentum expansion of the inverse propagator. This pregeometry seems to be a valid candidate for euclidean quantum gravity, without obstructions for analytic continuation to a Minkowski signature of the metric.
\end{abstract}


\maketitle

\medskip
\section{Introduction\label{sec: Introduction}}

In a modern view, quantum field theories are defined by functional integrals. For many models of particle physics one can employ an euclidean  functional integral and continue analytically to Minkowski signature. No well defined functional integral is known for a continuum formulation of quantum gravity based on the metric field. Such a functional integral would require the explicit specification of the microscopic or classical action. For a formulation in terms of the metric any action involving only a finite number of derivatives is problematic, however. The Einstein-Hilbert action with a cosmological constant involves up to two derivatives of the metric. The restriction to two derivatives is not compatible with a renormalizable theory. Furthermore, the euclidean action is not bounded from below, such that no corresponding euclidean functional integral can be defined. Adding invariants with up to four derivatives of the metric leads to a renormalizable theory~\cite{STE,FRATSE,AB}. Also the euclidean action can be bounded from below for appropriate signs of the couplings, such that an euclidean functional integral can be defined. The problem arises now from the presence of instabilities in the propagators in flat or weakly curved geometries. In flat space, ghosts or tachyons cannot be avoided for any finite polynomial momentum expansion of the inverse propagator beyond quadratic order in momenta. This implies that the addition of terms with a finite number of derivatives cannot cure the problem.

Quantum gravity based on the metric may be a non-perturbatively renormalizable quantum field theory. This is the case if the functional flow of a scale-dependent effective action~\cite{CWFE} admits an ultraviolet fixed point. Asymptotically safe quantum gravity~\cite{WEIN,MR,DPER,SOUM,RSAU, LAUR} is defined by choosing this ultraviolet fixed point for a definition of the short distance behavior of quantum gravity. The fixed point behavior permits to extrapolate this short distance behavior to arbitrarily short distances or high momenta. 

A fixed point corresponds to a scaling solution of an exact functional flow equation~\cite{CWFE}.
Such scaling solutions have been found for many approximated or ``truncated" forms of the scale-dependent effective action~\cite{BEGPP}. This makes it plausible that an ultraviolet fixed point indeed exists, rendering metric quantum gravity non- perturbatively renormalizable.

A functional integral can be defined formally as a solution of the exact flow equation. In practice, this is of little help if the solution gets too complicated.
This is presently the case for the functional renormalization group approach. The fixed point involves many invariants, and it is not clear what approximation is needed in order to obtain a well defined propagator for the graviton without ghost or tachyonic instabilities~\cite{PLCW}. This explains why so far no proposal for a simple classical action defining the functional integral for quantum gravity has been formulated in this approach.

One possibility to avoid these problems is a formulation of the functional integral in terms of fields different from the metric. This approach stays within the setting of quantum field theories. In analogy to quantum chromodynamics (QCD) for the strong interactions in particle physics, the short distance behavior - gluons and quarks in QCD - is described by fields different from the metric. The metric arises as a collective or  composite field in an effective low energy theory, in analogy to mesons and baryons in QCD. This type of quantum field theory with an emergent metric may be called ``pregeometry"~\cite{AK}. The most radical approach suggests to use only fermions as fundamental degrees of freedom~\citep{AK,AV,DS}. In spinor gravity a formulation with local Lorentz symmetry and diffeomorphism invariance has indeed been achieved~\cite{CWTS,CWSG1,CWSG2}. (For first formulations with global Lorentz symmetry see ref.~\cite{CWSGA,CWHEB}.) The vierbein and the metric have been constructed as composites of the fermions. While conceptually rather attractive, practical computations in this purely fermionic setting have not yet advanceed very much, due to the structure of the action containing only invariants with a rather high number of fermions.

The present work proposes a formulation of euclidean pregeometry as a $SO(4)$ - Yang-Mills theory. Besides fermions, which we do not discuss explicitly in the present paper, the degrees of freedom of pregeometry are the six gauge fields $A_{\mu}$ of the $SO(4)$ - gauge symmetry, and four additional vector fields $\tilde{e}_{\mu}$ which belong to the vector representation of $SO(4)$. This model has been discussed earlier~\cite{CWGG, CWFSI}, see also ref~\cite{RPER, Diakonov:2011im, Vladimirov:2012vw, Matsuzaki:2020qzf, Krasnov:2017epi} for related ideas. A diffeomorphism invariant action is based on the gauge invariant kinetic terms of the two sorts of vector fields $A_{\mu}$ and $\tilde{e}_{\mu}$,
\be 
\label{A0A}
\!\!S=\int_{x}\tilde{e}\Big{(}\frac{Z}{8}F_{\mu\nu mn} F_{\rho\sigma}{}^{mn}+\frac{1}{4}\tilde{U}_{\mu\nu m}\tilde{U}_{\rho\sigma}{}^{m}\Big{)} \tilde{e}_{p}{}^{\mu} \tilde{e}^{p\rho} \tilde{e}_{q}{}^{\nu} \tilde{e}^{q\sigma}.\!\!
\ee
Here $F_{\mu\nu mn} $ is the covariant field strength of the gauge boson $A_{\mu mn}=-A_{\mu nm}\:$ and $\tilde{U}_{\mu\nu}{}^{m}=D_{\mu}\tilde{e}_{\nu}{}^{m}$ the covariant derivative of the other vector field. With $\tilde{e}=det(\tilde{e}_{\mu}{}^{m})>0$ we can define $\tilde{e}_{m}{}^{\mu} $ by the relation $\tilde{e}_{m}{}^{\mu}\tilde{e}_{\mu}{}^{n}=\delta_{m}^{n}$ . 
The \mbox{$SO(4)$~- indices $m, n, p, q$} are raised and lowered with $\delta^{mn}$ or $\delta_{mn}$, such that their position can be regarded as arbitrary.
This action has a single dimensionless coupling $Z$ which is related to the gauge coupling.

In the present paper we demonstrate that the euclidean action involving the sum of the invariant kinetic terms for $A_{\mu}$ and $\tilde{e}_{\mu}$ is bounded from below. The euclidean functional integral is well defined for this type of pregeometry. The short distance propagators show neither ghosts nor tachyons. These stability properties may be expected since only terms with two derivatives of the fields are included in the effective action.

The metric is a composite field defined as a $SO(4)$ - invariant bilinear of the vector fields $\tilde{e}_{\mu}$. After a suitable rescaling the dimensionless vector fields $e_{\mu}$ play the role of generalized vierbeins. In contrast to Cartan's geometry~\cite{CAR} the covariant derivative of $e_{\mu}$ does not vanish, however. This distinguishes the present approach from many formulations where the gauge fields are associated to the spin connection which is a functional of the vierbein~\cite{CAR,UTI,KIB,SCI,HH,SHA,HCM,HEI,RP,PS,SSTZ,MMOY}. In our approach both the vierbein and the gauge fields contain propagating degrees of freedom beyond the ones contained in the metric. They are crucial for the different short distance behavior.
A formulation of generalized gravity in Minkowski space that is close to our setting has been developed in refs.~\cite{HS1, HS2, HS3, SVN, NARR, NIRRU}.
(For first functional renormalization group investigations of Cartan geometry see refs~\cite{DR2,HR,DR1}, and for alternative ideas on diffeomorphism invariant gauge theories cf. ref.~\cite{KR1,KR2}.)

The classical field equations have a solution with vanishing gauge fields $A_{\mu}$ and constant vierbeins $e_{\mu}$, leading for the composite metric $g_{\mu\nu}$ to a geometry of flat space, $g_{\mu\nu}=\delta_{\mu\nu}$. This solution entails a ``spontaneous symmetry breaking" of the $SO(4)$ - gauge symmetry. As a consequence of this ``generalized Higgs mechanism" the gauge bosons acquire a mass term and mix with the other vector field $\tilde{e}_{\mu}$. Linear combinations of both fields decouple from the effective low energy theory, which can be defined for momenta smaller than the gauge boson mass.

Field equations and effective low energy theory should be discussed in terms of the quantum effective action which includes all effects of quantum fluctuations. We will not perform here a computation of the quantum effective action. We rather make the assumption that the quantum fluctuations generate for the effective action all terms allowed by diffeomorphism and gauge symmetries. In particular, this includes a term linear in the derivatives of the gauge fields, and a cosmological constant. There are also additional terms with two derivatives of $A_{\mu}$. Terms with more than two derivatives will be neglected in the present paper.
 
The effective low energy theory is dominated by the terms with a small number of derivatives. In leading order this turns out to be general relativity for the metric, with an Einstein-Hilbert action and a possible cosmological constant that we assume to be very small, as required by observation. General relativity emerges naturally from the present formulation of pregeometry. 
This holds provided that the effective theory is stable in the sense that there are no tachyonic instabilities with mass terms of the order of the Planck mass or larger. We will see that this requires for the effective action the presence of invariants beyond those for the classical action~\eqref{A0A}.
Beyond the leading Einstein-Hilbert term the effective low energy theory contains terms with more than two derivatives for the metric. An expansion in the number of derivatives yields at fourth order an effective theory of the type of ref~\cite{STE}. One may therefore wonder what happens to the ghost instability in the graviton propagator for this fourth order gravity.
 
In our formulation of pregeometry the graviton fluctuation is a linear combination of fields contained in $A_{\mu}$ and $\tilde{e}_{\mu}$. As a result of the diagonalization of the propagator matrix we find that the dependence of the inverse graviton propagator on the squared momentum $q^{2}$ is given by a function 
 \be \label{IN1}
\begin{split}
&G_{\textup{grav}}^{-1}(q^{2})=\dfrac{m^{2}}{8}\bigg{\lbrace}(Z+1)q^{2}+m^{2}-M^{2} \\
&-\sqrt{\bigl[(Z-1)q^{2}+m^{2}-M^{2}\bigr]^{2}+4\dfrac{q^{2}}{m^{2}}\bigl(m^{2}-M^{2}\bigr)^{2}}\bigg{\rbrace}\;.
\end{split}
\ee
It involves three model parameters, namely $m^{2}$ which multiplies the kinetic term for the vierbein and determines the effective mass for the gauge bosons, the coupling $Z$ which multiplies the gauge boson kinetic term and equals the inverse squared gauge coupling, and $M^{2}$ which multiplies the term linear in derivatives and can be associated with the squared Planck mass.
Here $m^{2}$ only sets the overall mass scale, such that only two parameters $M^{2}/m^{2}$ and $Z$ specify a given model.
The absence of ghost or tachyonic instabilities is guaranteed for the parameter range
\be\label{AA}
0<M^{2}<m^{2}\;,\quad\quad 0<Z<Z_{c}\;,
\ee
with
\be\label{AB}
Z_{c}=\dfrac{M^{2}}{m^{2}}\bigg(1-\dfrac{M^{2}}{m^{2}}\bigg)^{-1}\;.
\ee
For this parameter range the graviton propagator is well behaved with a single pole at $q^{2}=0$ and without any instability. The ghost in fourth order gravity is an artefact of the truncated polynomial expansion~\cite{PLCW}.

The euclidean classical action can be continued analitically to Minkowski signature. The same holds for the graviton propagator~\eqref{IN1}.
So far we have not found any obstruction to analytic continuation. We may therefore consider our model of pregeometry as a proposal for a functional integral for quantum gravity. A recent investigation~\cite{CWPC} reveals that the field equations derived from the effective action with Minkowski signature lead to consistent and realistic cosmology if a scalar field is added. This includes an inflationary epoch and dynamical dark energy.

Our model of pregeometry is introduced in sect.~\ref{sec: Pregeometry}, where we also show that flat space solves the classical field equations. Sect.~\ref{sec: Generalized Higgs mechanism} discusses the generalized Higgs mechanism. It proceeds to a decomposition of the fluctuations around flat space according to different representations of its symmetry. This is the basis for the stability analysis of the classical theory which we address in sect.~\ref{sec: Stability of classical theory}. There we establish that no ghosts or tachyons are present in our model. In sect.~\ref{sec: Effective action and emergence of general relativity} we turn to the quantum effective action and the emergence of general relativity as an effective theory for small momenta and curvature. The graviton propagator is discussed in detail in sect.~\ref{sec: Graviton Propagator}. Together with a stable graviton our model of pregeometry also contains a stable massive spin-two particle. Sect.~\ref{sec: Stability of flat space} turns to the sector of scalar fields.
For suitable values of the couplings multiplying the invariants in the effective action we find that the scalar sector is stable as well.
Flat space is not the absolute minimum of the euclidean effective action in the space of all possible field configurations. It may, however, be the minimum in the space of solutions to the field equations. In sect.~\ref{sec: Cosmological constant} we add the effects of a cosmological constant. Sect.~\ref{sec: Conclusions and discussion} finally discusses our findings and possible extensions in various directions.

\section{Pregeometry\label{sec: Pregeometry}}

Similar to the electroweak or strong interactions our model is a non-abelian local gauge theory or Yang-Mills theory. For our euclidean setting the gauge group is $SO(4)$. The particularity as compared to the gauge theories in particle physics is the presence of an additional vector field $e_{\mu}{}^{m}(x)$ in the vector representation of $SO(4)$. We restrict the space of field configurations in the functional integral to $e=\det (e_{\mu}{}^{m})>0$, where $e_{\mu}{}^{m}$ is considered as a $4\times 4$ - matrix. This allows for the definition of the inverse matrix $e_{m}{}^{\mu}$, and the construction of diffeomorphism invariant kinetic terms for both the gauge fields $A_{\mu}$ and the vector fields $e_{\mu}$.

\subsection{Diffeomorphism invariant Yang-Mills theory\label{subsec: Diffeomorphism invariant Yang-Mills theory}}
\smallskip

Our starting point is a SO(4) - gauge theory with gauge fields
\al{
A_{\mu mn}=-A_{\mu nm}\; ,\qquad n,~m=0,\cdots ,3\;,
}
and field strength
\al{
\!\!\!\!F_{\mu\nu mn}=\p_\mu A_{\nu mn} -\p_\nu A_{\mu mn} + A_{\mu m}{}^p A_{\nu pn} -A_{\nu m}{}^p A_{\mu pn}.\!\!\!\!
\label{TDI2}}
The double-index $(m, n)=z$ labels the six gauge bosons $A_{\mu z}$ in the adjoint representation of $SO(4)$.
We further include vector fields $e_\mu{}^m$ in the vector representation of $SO(4)$. Accordingly, the covariant derivative reads
\al{
U_{\mu\nu}{}^m=D_\mu e_\nu{}^m=\p_\mu e_\nu{}^m -\Gamma_{\mu\nu}{}^\sigma e_\sigma{}^m + A_\mu{}^m{}_n e_\nu{}^n\,.
\label{field strength for vierbein}
}
These vector fields are restricted to obey
\al{
e=\det (e_\mu{}^m)\,>0\;,
\label{TDI07}}
where the components $e_\mu{}^m$ are considered as elements of a regular $4\times 4$ matrix.
This permits the definition of inverse vector fields $e_{m}{}^{\mu}\,$, with
\be\label{TDI1}
e_{\mu}{}^{m}e_{m}{}^{\nu}=\delta_{\mu}^{\nu}\,,\quad \quad e_{m}{}^{\mu}e_{\mu}{}^{n}=\delta_{m}^{n}\;.
\ee
While the vector field $\tilde{e}_{\mu}{}^{m} $ has the dimension mass usual for vector fields, the field $e_{\mu}{}^{m}=m^{-1}\tilde{e}_{\mu}{}^{m}$ is dimensionless. 

We introduce a composite metric $g_{\mu\nu}$ as a bilinear in the vector field and note that for $e\neq 0$ its inverse $g^{\mu\nu}$ exists,
\al{
&g_{\mu\nu} = e_\mu{}^m e_\nu{}^n \delta_{mn},&
&g^{\mu\nu} g_{\nu \rho} =\delta^\mu_\rho\,.
\label{eq(8)TDI}}
The connection $\Gamma_{\mu\nu}{}^\sigma$ in Eq.\,\eqref{field strength for vierbein} is the Levi-Civita connection formed with the composite metric $g_{\mu\nu}$. Inserting eq.~\eqref{eq(8)TDI} it can be expressed in terms of the vector field and its derivatives.
For $D_{\mu} e_{\nu}{}^{m}=0$ the vector field $e_{\mu}{}^{m}$ can play the role of the usual vierbein. We will not impose this constraint here such that the vector field contains additional degrees of freedom.

The field strengh~\eqref{TDI2} and the covariant derivative~\eqref{field strength for vierbein} transform as tensors. These tensors can be used as basic building blocks for the construction of invariant kinetic terms for the gauge fields $A_{\mu mn}$ and the vector field $e_{\mu}{}^{m}$. We define a diffeomorphism invariant action
\al{
S=\int\df^4x\,e ( L_F + L_U)\,,
\label{starting action}
}
with  gauge field kinetic term
\al{
L_F = \frac{Z}{8}F_{\mu\nu mn} F^{\mu\nu mn}\,,
}
and  vector field kinetic term
\al{
L_U = \frac{m^2}{4} U_{\mu\nu m}U^{\mu\nu m}\,.
}
The mass $m$ arises uniquely from the rescaling of $\tilde{e}_{\mu}{}^{m}$ and should not be regarded as a free parameter.
World (space-time) indices $\mu$, $\nu$ are raised and lowered with $g^{\mu\nu}$ and $g_{\mu\nu}$ given by eq.~\eqref{eq(8)TDI}, while ``Lorentz indices" $m$, $n$ are raised and lowered with $\delta^{mn}$ and $\delta_{mn}$.
We can convert Lorentz indices to world indices by multiplication with $e_{\mu}{}^{m}$ or $e_{m}{}^{\mu}$,
\be
U_{\mu\nu\rho}=U_{\mu\nu}{}^m e_{\rho m}\;.
\ee
For $U_{\mu\nu}{}^{ m}\neq 0$ the covariant derivatives commute with the raising and lowering of world indices or Lorentz indices, but not with the conversion from world to Lorentz indices,
\be\label{TDI3}
D_{\sigma}U_{\mu\nu\rho}=\left( D_{\sigma}U_{\mu\nu}{}^{m}\right)e_{\rho m}+ U_{\mu\nu}{}^{m}U_{\sigma\rho m}\;.
\ee
Because of their analogous role to the vierbein in Cartan's geometry~\cite{CAR} we will call the vector fields $e_{\mu}{}^{m}$ ``generalized vierbeins", or simply ``vierbeins". 

Two different connections are present in our setting, the gauge connection $A_{\mu mn}$ and the geometric connection $\Gamma_{\mu\nu}{}^{\rho}$. The tangent bundle and the gauge bundle are, a priori, not related. For example, one can have the metric corresponding to a sphere and nevertheless vanishing gauge fields, $A_{\mu mn}=0$. This permits, for example, to have massless fermions (zero eigenvalues of the fermionic kinetic operator) on a sphere \cite{CWGG}. A relation between the connections can be formulated by the identity
\be
U_{\mu\nu\rho}=\omega_{\mu\nu\rho} - A_{\mu\nu\rho}\,,
\ee
with $\omega_{\mu\nu\rho}$ related to the ``spin connection" formed from the vector field
\begin{align}\label{15AA}
\omega_{\mu\nu\rho}
&=\frac{1}{2}\bigg\{ e_{\mu m}\left( \p_\rho e_\nu{}^m -\p_\nu e_\rho{}^m\right) + e_{\nu m}\left( \p_\rho e_\mu{}^m -\p_\mu e_\rho{}^m \right) \nn \\
&\quad +e_{\rho m} \left( \p_\mu e_\nu{}^m -\p_\nu e_\mu{}^m \right) \bigg\}=e_{\nu}{}^m e_\rho{}^n \omega_{\mu mn} \;.
\end{align}
This expresses the Levi-Civita connection in terms of the vierbein by
\be\label{16A}
\omega_{\mu\nu\rho}=\Gamma_{\mu\rho}{}^\sigma g_{\sigma\nu} -e_{\nu m}\p_\mu e_\rho{}^m
\,.
\ee
We observe the antisymmetry in the last two indices
\be\label{16AA}
\omega_{\mu\rho\nu}=-\omega_{\mu\nu\rho}\;,\;\; A_{\mu\rho\nu}=-A_{\mu\nu\rho}\;,\;\; U_{\mu\rho\nu}=-U_{\mu\nu\rho}\;.
\ee

For a covariantly conserved vector field, $U_{\mu\nu\rho}=0$, the gauge field equals the spin connection formed from the vierbein. This is the setting of Cartan's geometry~\cite{CAR}. One can define the torsion tensor
\be\label{19A}
T_{\mu\nu\rho}=A_{\rho\mu\nu}-A_{\nu\mu\rho}-e_{\mu m}(\partial_{\nu}e_{\rho}{}^{m}-\partial_{\rho}e_{\nu}{}^{m})\;.
\ee
For Minkowski signature and identified gauge and tangent bundles this is the object investigated in refs.~\cite{HS1, HS2, HS3, SVN, NARR, NIRRU}.
The covariant vierbein derivative can be expressed in terms of the torsion as
\be\label{19B}
U_{\mu\nu\rho}=\frac{1}{2}(T_{\mu\nu\rho}-T_{\nu\rho\mu}+T_{\rho\nu\mu}).
\ee

\subsection{Boundedness of the action and euclidean \\functional integral\label{subsec: Boundedness of the action and euclidean functional integral}}
\smallskip

The action~\eqref{starting action} is bounded from below, $S\geqslant 0$, as long as $m^{2}\geqslant 0$, $Z\geqslant 0$. In order to see this, we first show that the composite metric has only positive eigenvalues. Since $g_{\mu\nu}$ is a real symmetric matrix, it can be diagonalized by an orthogonal transformation $O$,
\begin{align}\label{TDI K11}
\tilde{g}_{\rho\sigma}=&O_{\rho}{}^{\mu}O_{\sigma}{}^{\nu}g_{\mu\nu}=O_{\rho}{}^{\mu}e_{\mu}{}^{m}O_{\sigma}{}^{\nu}e_{\nu}{}^{n}\delta_{mn}\nn \\
=&\tilde{e}_{\rho}{}^{m}\tilde{e}_{\sigma}{}^{n}\delta_{mn}\;,
\end{align}
with $\tilde{e}_{\rho}{}^{m}=O_{\rho}{}^{\mu}e_{\mu}{}^{m}$. The diagonal elements are indeed positive (no sum over $\rho$)
\be\label{TDI K12}
\tilde{g}_{\rho\rho}=\sum_{m}(\tilde{e}_{\rho}{}^{m})^{2}>0\;.
\ee
For $e>0$ one can exclude vanishing eigenvalues $\tilde{g}_{\rho \rho}$. 

We next write for $z=(m,n)$
\be\label{TDI K13}
\!\!\!\!F_{\mu\nu mn}=F_{\alpha z}\; , \quad\quad F_{\mu\nu m n} F^{\mu\nu m n }=\sum_{z} F_{\alpha z} G^{\alpha\beta} F_{\beta z }\;,
\ee
with $\alpha=(\mu ,\nu)$, $\beta=(\rho,\sigma)$ double indices, and real symmetric matrix
\be\label{TDI K14}
G^{\alpha\beta}=g^{\mu\rho}g^{\nu\sigma}=G^{\beta\alpha}\;.
\ee
The expression $L_{F}$ is positive semidefinite if $G^{\alpha\beta}$ has only positive eigenvalues. This is indeed the case, as can be seen by diagonalizing $G^{\alpha\beta}$ by an orthogonal transformation. We can chose the required orthogonal matrix as the direct product of the matrices that diagonalize $g^{\mu\nu}$, resulting in $\tilde{G}^{\alpha\alpha}=\tilde{g}^{\mu\mu}\tilde{g}^{\nu\nu}>0$. The same argument shows $L_{U}\geqslant 0$, with $z=m$ in this case. With both $L_{U}$ and $L_{F}$ positive semidefinite and $e>0$, the action is indeed positive semidefinite, $S\geq 0$.
The action vanishes precisely if both $L_{F}$ and $L_{U}$ vanish.

The positivity of the classical action~\eqref{starting action} for $Z>0\, $, $m\2>0$ can also be seen from the identities $F_{\mu\nu mn}F^{\mu\nu mn}=F_{pqmn}F^{pqmn}$, $U_{\mu\nu m}U^{\mu\nu m}=U_{pqm}U^{pqm}$, where $F_{pqmn}=e_{p}{}^{\mu}e_{q}{}^{\nu}F_{\mu\nu mn}$. Transmutations of world indices $\mu ,\nu$ to Lorentz indices $p,q$ are achieved by multiplication with the vierbein or inverse vierbein. Since index contractions of Lorentz indices are simply index summations, both $L_{F}$ and $L_{U}$ are squares.

With positive semidefinite classical action $S$ an euclidean functional integral can be defined by integrating over unconstrained gauge fields $A_{\mu}$ and vector fields $e_{\mu}$ obeying the constraint~\eqref{TDI07}. The constraint $e>0$ is invariant under local $SO(4)$ - gauge transformations and diffeomorphisms. In presence of these two local symmetries the action needs a regularization, either by the usual gauge fixing procedure for a continuous formulation, or perhaps by a lattice formulation that still has to be found. We consider the action~\eqref{starting action} as a well defined and simple starting point for an investigation of quantum fluctuactions in euclidean gravity. The analytic continuation to a Minkowski signature will be discussed in sect.~\ref{sec: Graviton Propagator}.

Configurations with a constant vierbein and vanishing gauge fields include the``flat space configuration"
\be\label{F13}
e_{\mu}{}^{m}=\delta_{\mu}^{m}\;,\quad g_{\mu\nu}=\delta_{\mu\nu}\;,\quad A_{\mu mn}=0\;.
\ee
These configurations yield $U_{\mu\nu}{}^{m}=0$ and are therefore minima of the action~\eqref{starting action}. They are solutions of the corresponding field equations that we discuss in more detail in sect.~\ref{sec: Effective action and emergence of general relativity}.

\section{Generalized Higgs mechanism\label{sec: Generalized Higgs mechanism}}

In order to understand the physical content of our model at short distances we decompose the fluctuations of the vector field and the gauge fields around flat space and zero gauge fields. We work in euclidean flat space. Analytic continuation to Minkowski space  can be performed at the end. For sufficiently short distances one can neglect any effects of a given geometry, as encoded in $e_{\mu}{}^{m}$ or $g_{\mu\nu}$, or of a given gauge field configuration $A_{\mu m n}$. For the short distance limit the flat space configuration $e_{\mu}{}^{m}=\delta_{\mu}^{m}$, $A_{\mu m n}=0$ is always an appropriate starting point for a field-expansion.

The decomposition involves different representations of the ``euclidean Lorentz group" $SO(4)$ that cannot mix in quadratic order. Mixing is observed, however, between modes belonging to the same representation. This renders the discussion of stability somewhat complex. In the present section we discuss the general structure of the mixing and focus subsequently on the graviton propagator and the scalar propagator in the next section. We will find a stable graviton propagator without ghosts or tachyons. 
The discussion will be extended in sect.~\ref{sec: Effective action and emergence of general relativity} to the quantum effective action. The graviton propagator for this extended setting is discussed in sect.~\ref{sec: Graviton Propagator}, and scalars are addressed in sect.~\ref{sec: Stability of flat space}.

Due to spontaneous symmetry breaking by the non-vanishing value of the vierbein $e_{\mu}{}^{m}=\delta_{\mu}^{m}$ mass terms are generated for the gauge bosons. This``generalized Higgs mechanism" differs from the usual Higgs mechanism since the agent of spontaneous symmetry breaking is a vector field and not a scalar field. The mode decomposition reveals the detailed structure of the generalized Higgs mechanism by focusing on the effects for the propagators of the various fluctuation modes. The mode mixing is a characteristic feature of this generalized Higgs mechanism.

\subsection{Fluctuations in flat space\label{subsec: Fluctuations in flat space}}
\smallskip

The field equations derived from the action \eqref{starting action} admit as a solution flat space, $e_\mu{}^m=\delta_\mu^m$, $g_{\mu\nu}=\delta_{\mu\nu}$, with vanishing gauge fields $A_{\mu mn}=0$.
This solution will be found to be stable, as may be expected since it corresponds to the minimum of the action. For a more detailed investigation of the stability for various modes we expand in quadratic order in $H_{\mu\nu}$ and $A_{\mu mn}$,
\al{
e_\mu{}^m=\delta_\mu^m + \frac{1}{2} H_{\mu\nu}\delta^{\nu m}\,.
}
The standard kinetic term for the transverse gauge fields reads in momentum space
\al{
\int_x\, e L_F= \frac{Z}{4} \int_q A_{\mu mn}(-q) \left( q^2 \delta^{\mu\nu} -q^\mu q^\nu\right) A_\nu{}^{mn}(q)\,.
\label{kinetic Lagrangian for LL gauge field}
}
This dominates the high momentum behavior and ensures stability in this range. For the discussion of fluctuations world indices are raised and lowered with $\delta^{\mu\nu}$ and $\delta_{\mu\nu}$.

In quadratic order the covariant kinetic term $L_{U}$ for the vierbein contains three pieces
$L_U=L_U^{(1)}+ L_U^{(2)}+L_U^{(3)}$. 
The first piece is a kinetic term which reads, with $\p^2=\p^\rho \p_\rho$,
\al{
&\int_x\,e  L_U^{(1)}=-\frac{m^2}{16}\int_x \bigg\{ H^{(\text{A})}_{\mu\nu} \p^2 H^{(\text{A})\mu\nu} \\
&+2H_{\mu\nu}^{(\text{S})}(\p^2 \delta_\rho^\nu -\p^\nu \p_\rho) H^{(\text{S})}{}^{\rho\mu} 
+4H^{(\text{A})}_{\mu\nu} \p^\nu \p_\rho H^{(\text{A})}{}^{\rho\mu} \bigg\}\,. \nn
\label{kinetic term of H}
}
Here $H_{\mu\nu}^{(\text{S})}=(H_{\mu\nu}+H_{\nu\mu})/2$ and $H_{\mu\nu}^{(\text{A})}=(H_{\mu\nu}-H_{\nu\mu})/2$ are the symmetric and antisymmetric parts of the fluctuation $H_{\mu\nu}$.
The second piece acts as a mass term for the gauge bosons. Without mixing effects (see below), it would generate an equal mass for all gauge bosons
\al{
eL_U^{(2)}=\frac{m^2}{4} A_{\mu mn} A^{\mu mn}\,.
}
Finally, the third piece is a type of source term for the gauge bosons,
\al{
&e L_U^{(3)}=-J_{\mu\nu\rho}A^{\mu\nu\rho}\,,
\label{TDI eq: 15}}
with
\al{
& J_{\mu\nu\rho} =\frac{m^2}{2} \left( \p_\mu H_{\nu\rho}^{(\text{A})} +\p_\rho H_{\mu\nu}^{(\text{S})} -\p_\nu H_{\mu\rho}^{(\text{S})} \right)\,.
\label{TDI eq: 16}}

The field value $e_\mu{}^m=\delta_\mu^m$ breaks the $SO(4)$ - gauge symmetry spontaneously. Similarly to the Higgs mechanism, this spontaneous symmetry breaking generates a mass term for the gauge bosons.
In contrast to the Higgs mechanism the field $e_\mu{}^m$ is a vector, not a scalar.
Flat space preserves a global $SO(4)$ -``Lorentz" symmetry of simultaneous $SO(4)$ - gauge rotations and $SO(4)$ - coordinate transformations. It is this residual symmetry that we employ for the mode expansion.

\subsection{Decomposition of vierbein fluctuations\label{subsec: Decomposition of vierbein fluctuations}}
\smallskip

For an analysis of the kinetic term \eqref{kinetic term of H} for the vierbein we employ the decomposition  which orders the scalar fluctuations into a physical and a gauge degree of freedom~\cite{CWMF},
\al{
H_{\mu\nu}^{(\text{S})}&= t_{\mu\nu} + \p_\mu \kappa_\nu +\p_\nu \kappa_\mu \nn \\
&\quad +\frac{1}{3}\left( \delta_{\mu\nu} -\frac{\p_\mu \p_\nu}{\p^2} \right)\sigma +\frac{\p_\mu \p_\nu}{\p^2}u\;,
\label{40AA}}
and
\be\label{TDI Eq: 17}
H_{\mu\nu}^{(\text{A})}= b_{\mu\nu}  + \p_\mu \gamma_\nu -  \p_\nu \gamma_\mu\;.
\ee
Here $t_{\mu\nu}$, $b_{\mu\nu}$, $\kappa_{\mu}$ and $\gamma_{\mu}$ are all transversal,
\al{
&\p^\mu t_{\mu\nu} =\p^\nu t_{\mu\nu} = \p^\mu b_{\mu\nu} =\p^\nu b_{\mu \nu}=0\,, \nn \\
&\p^\mu \kappa_\mu =\p^\mu \gamma_\mu =0\,,
\qquad
t_{\mu}{}^\mu=0\,.
}
The kinetic piece reads in momentum space
\al{
\int_x\,e  L_U^{(1)}
&=\frac{m^2}{16}\int_q \bigg\{ 2 t_{\mu\nu}(-q)q^2 t^{\mu\nu}(q) +b_{\mu\nu}(-q) q^2 b^{\mu\nu}(q)\nn \\
&+ 2 \left( \kappa_\mu(-q) -\gamma_\mu(-q)\right)q^4 \left( \kappa^\mu(q) -\gamma^\mu(q)\right) \nn \\
&\quad
+\frac{2}{3}\sigma(-q)q^2 \sigma(q)\bigg\}\,.
\label{TDI Eq: 19}}
The factor $q^{4}$ for $\kappa_{\mu}-\gamma_{\mu}$ is a consequence of the particular normalization of these fields. Indeed, we can absorb a factor $4q^2$ in a dimensionless normalization of $\kappa_\mu$ and $\gamma_\mu$. All kinetic terms are then positive and increase $\sim q^2$. The kinetic term $L_U^{(1)}$ does not involve the combination $\kappa_\mu+\gamma_\mu$ and the scalar $u$.
Similarly, $L_F$ does not involve the longitudinal components $\p^\mu A_{\mu mn}$ of the gauge fields.
These fields correspond to gauge degrees of freedom.

\subsection{Decomposition of gauge field fluctuations\label{subsec: Decomposition of gauge field fluctuations}}
\smallskip

In flat space we need no longer to distinguish between world and Lorentz indices. Covariant derivatives become partial derivatives. We decompose the linear gauge fields $A_{\mu\nu\rho}$ into transversal modes $B_{\mu\nu\rho}$ and longitudinal modes $L_{\nu\rho}$,
\begin{equation}\label{DC01}
A_{\mu\nu\rho}=B_{\mu\nu\rho}+\partial_{\mu}L_{\nu\rho}\,,\quad \partial^{\mu}B_{\mu\nu\rho}=0\;.
\end{equation}
Using the projector
\begin{equation}\label{DC02}
P_{\mu}^{\nu}=\delta_{\mu}^{\nu}-\dfrac{\partial_{\mu}\partial^{\nu}}{\partial^{2}}\,,\quad\partial^{\mu}P_{\mu}^{\nu}=0\,,\quad P_{\mu}^{\nu}\partial_{\nu}=0\;,
\end{equation}
we can write
\begin{equation}\label{DC03}
B_{\mu\nu\rho}=P_{\mu}^{\sigma}A_{\sigma\nu\rho}=P_{\mu}^{\sigma}B_{\sigma\nu\rho}\;.
\end{equation}
The transversal fluctuations can be decomposed as
\be\label{DC04}
\begin{split}
B_{\mu\nu\rho}=\dfrac{1}{4}\varepsilon_{\nu\rho}^{\quad\sigma\tau}(P_{\mu\sigma}v_{\tau}-P_{\mu\tau}v_{\sigma})\quad\quad  \\
+\dfrac{1}{3}(P_{\mu\nu}w_{\rho}-P_{\mu\rho}w_{\nu})+D_{\mu\nu\rho}\;,
\end{split}
\ee
where $v_{\mu}$ and $w_{\mu}$ are (reducible) four-vectors. The decomposition of the remaining part $D_{\mu\nu\rho}$ into irreducible representations is given by
\be\label{DC04a}
D_{\mu\nu\rho}=\dfrac{1}{2}(\partial_{\nu}E_{\mu\rho}-\partial_{\rho}E_{\mu\nu})+C_{\mu\nu\rho}\;.
\ee

The transversal traceless symmetric tensor $E_{\mu\nu}$ obeys
\begin{equation}\label{DC05}
\partial^{\mu}E_{\mu\nu}=0\,,\quad\delta^{\mu\nu}E_{\mu\nu}=0\,,\quad E_{\mu\nu}=E_{\nu\mu}\;,
\end{equation}
and $C_{\mu\nu\rho}$ is subject to the constraints
\begin{align}\label{DC06}
C_{\mu\nu\rho}=-C_{\mu\rho\nu}\,,\quad \partial^{\mu}C_{\mu\nu\rho}=0\,,\quad\varepsilon^{\mu\nu\rho\sigma}C_{\mu\nu\rho}=0\,,\nn\\
\delta^{\mu\nu}C_{\mu\nu\rho}=\delta^{\mu\rho}C_{\mu\nu\rho}=0\,,\quad \tilde{P}_{\sigma\tau}^{\quad\nu\rho}C_{\mu\nu\rho}=0\;.
\end{align}
We have introduced here a further projector
\begin{equation}\label{DC07}
\tilde{P}_{\sigma\tau}^{\quad\nu\rho}=\dfrac{1}{2\partial^{2}}(\partial_{\sigma}\partial^{\nu}\delta_{\tau}^{\rho}-\partial_{\tau}\partial^{\nu}\delta_{\sigma}^{\rho}-\partial_{\sigma}\partial^{\rho}\delta_{\tau}^{\nu}+\partial_{\tau}\partial^{\rho}\delta_{\sigma}^{\nu})\;,
\end{equation}
which obeys
\begin{align}\label{DC08}
\tilde{P}_{\sigma\tau}^{\quad\alpha\beta}&\tilde{P}_{\alpha\beta}^{\quad\nu\rho}=\tilde{P}_{\sigma\tau}^{\quad\nu\rho}\,,\quad \tilde{P}_{\nu\rho}^{\quad\nu\rho}=3\,,\nn\\ 
 \partial^{\tau}&\tilde{P}_{\sigma\tau}^{\quad\nu\rho} =\dfrac{1}{2}(\partial^{\rho}\delta_{\sigma}^{\nu}-\partial^{\nu}\delta_{\sigma}^{\rho})\;.
\end{align}
This results in the identities
\be\label{DC09}
\tilde{P}_{\sigma\tau}^{\quad\nu\rho} D_{\mu\nu\rho}=(\partial_{\sigma}E_{\mu\tau}-\partial_{\tau}E_{\mu\sigma})\,,
\ee
and
\be
\tilde{P}_{\sigma\tau}^{\quad\nu\rho} (\partial_{\nu}E_{\mu\rho}-\partial_{\rho}E_{\mu\nu})=(\partial_{\sigma}E_{\mu\tau}-\partial_{\tau}E_{\mu\sigma})\;.
\ee
We can therefore see $C_{\mu\nu\rho}$ as a projection,
\be
C_{\mu\nu\rho}=D_{\mu\nu\rho}-\tilde{P}_{\nu\rho}^{\quad\sigma\tau}D_{\mu\sigma\tau}\;.
\ee

Out of the 24 components $A_{\mu\nu\rho}$ six degrees of freedom $L_{\nu\rho}=-L_{\rho\nu}$ are longitudinal degrees of freedom. Thus there are 18 transversal modes $B_{\mu\nu\rho}$. The four modes $v_{\mu}$ correspond to the totally antisymmetric part $A_{[\mu\nu\rho]}$, while the four modes $w_{\rho}$ account for the trace $A_{\mu\nu\rho}\delta^{\mu\nu}$. There remain 10 modes for $D_{\mu\nu\rho}$. The projector $\tilde{P}$ eliminates half of them, and the traceless transversal symmetric tensor $E_{\mu\nu}$ accounts indeed for five modes. The other five modes correspond to $C_{\mu\nu\rho}=-C_{\mu\rho\nu}$, which indeed is subject to 19 constraints. The vectors $v_{\mu}$ and $w_{\mu}$ can be decomposed into transversal vectors and scalars
\begin{align}\label{DC10}
&v_{\mu}=v_{\mu}^{(t)}+\partial_{\mu}\tilde{v}\,,\quad w_{\mu}=w_{\mu}^{(t)}+\partial_{\mu}\tilde{w}\,,\nn\\ &\partial^{\mu}v_{\mu}^{(t)}=0\,,\quad \partial^{\mu}w_{\mu}^{(t)}=0\;.
\end{align}
The irreducible representations of the transversal modes $B_{\mu\nu\rho}$ are $2\times 5 + 2\times 3+2\times 1$. In particular, the scalar part of $B_{\mu\nu\rho}$ reads
\be\label{DC11}
B_{\mu\nu\rho}^{(s)}=\dfrac{1}{2}\varepsilon_{\mu\nu\rho}^{\quad\;\,\tau}\partial_{\tau}\tilde{v}+\dfrac{1}{3}(\delta_{\mu\nu}\partial_{\rho}\tilde{w}-\delta_{\mu\rho}\partial_{\nu}\tilde{w})\;.
\ee

Finally, the six longitudinal modes are two triplets
\be\label{DC12}
L_{\nu\rho}=M_{\nu\rho}+\partial_{\nu}l_{\rho}-\partial_{\rho}l_{\nu}\;,
\ee
with
\be\label{DC13}
\partial^{\nu}M_{\nu\rho}=\partial^{\rho}M_{\nu\rho}=0\,,\quad\partial^{\nu}l_{\nu}=0\;.
\ee
We summarize the various modes in table~\ref{Ta1}, with a slight abuse of notation for the physical modes in the vector sector.
\begin{table}[h]
\renewcommand\arraystretch{2}
\renewcommand\tabcolsep{6pt}
\begin{tabular}{|c|c|c|c|c|}
\hline
\multicolumn{2}{|c|}{Physical modes} & \multicolumn{2}{c|}{Gauge modes} & Type  \\
\hline
vierbein & gauge field & vierbein &
gauge field & \\
\hline
$t_{\mu\nu}$ & $E_{\mu\nu}$, $C_{\mu\nu\rho}$ &   &   & $T$, $5$ \\
\hline
$\sigma$ & $\tilde{v}$, $\tilde{w}$  & $u$ &   & $S$, $1$ \\
\hline
$\gamma_{\mu}$, $b_{\mu\nu}$ &$v_{\mu}^{(t)}$, $w_{\mu}^{(t)}$& $\kappa_{\mu}$ & $M_{\mu\nu}$, $l_{\mu}$ & $V$, $3$\\
\hline
\end{tabular}
\caption{Physical and gauge modes in the expansion of vierbein and gauge fields in flat space. We indicate tensors ($T$), transverse vectors ($V$) and scalars $S$, with the corresponding number of independent corresponds. In the vector sector the physical modes are actually linear combinations, given by the ones appearing in eqs~\eqref{DC26},~\eqref{DC27}. Setting $\kappa_{\mu}=l_{\mu}=M_{\mu\nu}=0$ we can interprete $\gamma_{\mu}$ and $b_{\mu\nu}$ as substitute for these linear combinations.}\label{Ta1}
\end{table}

\subsection{Generalized Higgs mechanism\label{subsec: Generalized Higgs mechanism}}
\smallskip

In contrast to spontaneous symmetry breaking by a scalar field the expectation value of a vector field leads to a mixing of the gauge field fluctuations and the vierbein fluctuations. This particular feature of the generalized Higgs mechanism is due to the ``source term" $L_{U}^{(3)}$. It is our aim to separate the various physical modes. For this purpose we have to diagonalize the inverse propagator which corresponds to the second functional derivative of the action. In other words, we have to find the field combinations for which the quadratic expansion of the action in the fluctuations becomes diagonal. We end this section by writing the quadratic expansion of the action in terms of the various fields appearing in our decomposition. In the next section we will proceed to the diagonalization.

The source term $L_{U}^{(3)}$ mixes the gauge fields $A_{\mu\nu\rho}$ and the vierbein fluctuations $H_{\mu\nu}$. In quadratic order one finds from eqs.~\eqref{TDI eq: 15}\eqref{TDI eq: 16}
\be\label{DC14}
\begin{split}
\int_{x}\!\overline{e}L_{U}^{(3)}\!=-\dfrac{m^{2}}{4}\!\!\!\int_{x}\!\!\!A^{\mu\nu\rho}\Bigl(\partial_{\mu}H_{\nu\rho}^{(A)}\!+\!\partial_{\rho}H_{\mu\nu}^{(S)}\!-\!\partial_{\nu}H_{\mu\rho}^{(S)}\Bigr)\\
=\dfrac{m^{2}}{4}\int_{x}\biggl{\lbrace}L^{\nu\rho}\Bigl(\partial^{2}H_{\nu\rho}^{(A)}+\partial_{\rho}\partial^{\mu}H_{\mu\nu}^{(S)}-\partial_{\nu}\partial^{\mu}H_{\mu\rho}^{(S)}\Bigr) \\
+2\partial_{\rho}B^{\mu\nu\rho}H_{\mu\nu}^{(S)}\biggr{\rbrace}=\dfrac{m^{2}}{4}\int_{x}(Y_{l}+Y_{t})\;.
\end{split}
\ee
The longitudinal part $Y_{l}$ concerns the mixing effects for the longitudinal gauge bosons. By use of the decomposition~\eqref{TDI Eq: 17},\eqref{DC12}, this results in
\be\label{DC15}
Y_{l}=M^{\nu\rho}\partial^{2}b_{\nu\rho}+2l^{\mu}\partial^{4}(\kappa_{\mu}-\gamma_{\mu})\;.
\ee

The transversal part $Y_{t}$ accounts for the mixing of the transversal gauge bosons. For its evaluation we decompose the tensor~$\partial_{\rho}B^{\mu\nu\rho}$ into its traceless symmetric, antisymmetric, and (modified) trace parts
\be\label{DC16}
\partial_{\rho}B^{\mu\nu\rho}=\tilde{B}^{(S)\mu\nu}+\tilde{B}^{(A)\mu\nu}+\dfrac{1}{3}P^{\mu\nu}\tilde{b}\;,
\ee
with
\begin{align}\label{DC17}
\!\!\!\!\tilde{B}^{(S)\mu\nu}\!\!=\!\tilde{B}^{(S)\nu\mu}&,\quad\!\! \tilde{B}^{(S)\mu\nu}\delta_{\mu\nu}\!\!=\!0,\quad\!\!\tilde{B}^{(A)\mu\nu}\!=\!-\tilde{B}^{(A)\nu\mu}\,,\!\!\!\!\nn\\
\partial_{\mu}&\tilde{B}^{(S)\mu\nu}+\partial_{\mu}\tilde{B}^{(A)\mu\nu}=0\;.\!\!\!\!
\end{align}
Comparing with the expansion~\eqref{DC04} we identify
\ba\label{DC18}
&\tilde{B}^{(S)\mu\nu}=-\dfrac{1}{2}\partial^{2}E^{\mu\nu}\,,\quad \tilde{B}^{(A)\mu\nu}=\dfrac{1}{2}\varepsilon^{\mu\nu\rho\tau}\partial_{\rho}v_{\tau}^{(t)}\,,\nn\\ 
&\quad\quad\tilde{b}=\partial^{2}\tilde{w}\;.
\end{align}
The components $C_{\mu\nu\rho}$ do not appear due to the identity
\be\label{DC19}
\partial^{\rho}C_{\mu\nu\rho}=0\;,
\ee
which follows from eqs.~\eqref{DC08}~\eqref{DC09}. 

Inserting also the expansion~\eqref{TDI Eq: 17} for $H_{\mu\nu}^{(S)}$, the transversal part of eq.~\eqref{DC14} reads
\be\label{DC20}
\!\!\!\! Y_{t}\!=\!\Bigl(\!2\tilde{B}^{(S)\mu\nu}\!+\dfrac{2}{3}P^{\mu\nu}\tilde{b}\Bigr)H_{\mu\nu}^{(S)}=\!-E^{\mu\nu}\partial^{2}t_{\mu\nu}\!+\dfrac{2}{3}\tilde{w}\partial^{2}\sigma\,.\!\!\!
\ee
We observe that the twelve transversal gauge field fluctuations $v^{(t)}$, $w^{(t)}$, $\tilde{v}$, $C_{\mu\nu\rho}$ do not appear in $L_{ U}^{(3)}$. They do not take part in the mixing and simply acquire a mass from $L_{U}^{(2)}$.
Also the four gauge modes $\kappa_{\mu}+\gamma_{\mu}$ and $u$ are absent.

For a discussion of the inverse propagator we also have to decompose the gauge boson mass term~$L_{U}^{(2)}$,
\begin{align}\label{DC21}
&L_{U}^{(2)}=\dfrac{m^{2}}{4}A^{\mu\nu\rho}A_{\mu\nu\rho}=\dfrac{m^{2}}{4}(B^{\mu\nu\rho}B_{\mu\nu\rho}-L^{\nu\rho}\partial^{2}L_{\nu\rho})\nn\\
=&\dfrac{m^{2}}{4}\Bigl{\lbrace}-\dfrac{1}{2}E^{\nu\rho}\partial^{2}E_{\nu\rho}+C^{\mu\nu\rho}C_{\mu\nu\rho}+v^{(t)\mu}v_{\mu}^{(t)}-\dfrac{3}{2}\tilde{v}\partial^{2}\tilde{v}\nn\\
&+\dfrac{4}{9}w^{(t)\mu}w_{\mu}^{(t)}-\dfrac{2}{3}\tilde{w}\partial^{2}\tilde{w}-M^{\nu\rho}\partial^{2}M_{\nu\rho}+2l^{\mu}\partial^{4}l_{\mu}\Bigr{\rbrace}\,.\! \!\! \!
\end{align}
The corresponding expression for $L_{U}^{(1)}$ is given by eq.~\eqref{TDI Eq: 19} 
\be\label{DC22}  
\begin{split}
L_{U}^{(1)}=&\dfrac{m^{2}}{4}\Bigl{\lbrace}-\dfrac{1}{2}t^{\mu\nu}\partial^{2}t_{\mu\nu}-\dfrac{1}{4}b^{\mu\nu}\partial^{2}b_{\mu\nu}\\&+\dfrac{1}{2}(\kappa^{\mu}-\gamma^{\mu})\partial^{4}(\kappa_{\mu}-\gamma_{\mu}) 
-\dfrac{1}{6}\sigma\partial^{2}\sigma\Bigr{\rbrace}\;.
\end{split}
\ee

\subsection{Mode mixing\label{subsec: Mode mixing}}
\smallskip

Taking things together, $L_{U}$ involves a couple of independent pieces
\be\label{DC23}
L_{U}=\dfrac{m^{2}}{4}\bigl{\lbrace} L_{tE}+L_{\sigma\tilde{w}}+L_{bM}+L_{\kappa\gamma l}+L_{m}'\bigr{\rbrace}\;.
\ee
They can be diagonalized separately.
For the transversal traceless symmetric tensors $t$ and $E$ one has
\be\label{DC24}
L_{tE}=-\dfrac{1}{2}(t^{\mu\nu}+E^{\mu\nu})\partial^{2}(t_{\mu\nu}+E_{\mu\nu})\;,
\ee
while in the scalar sector $\sigma$ and $\tilde{w}$ are connected
\be\label{DC25}
L_{\sigma\tilde{w}}=-\dfrac{1}{6}(\sigma-2\tilde{w})\partial^{2}(\sigma-2\tilde{w})\;.
\ee
The sector for the longitudinal gauge bosons involves
\be\label{DC26}
L_{b M}=-\dfrac{1}{4}(b^{\mu\nu}-2M^{\mu\nu})\partial^{2}(b_{\mu\nu}-2M_{\mu\nu})\;,
\ee
and
\be\label{DC27}
L_{\kappa\gamma l}=\dfrac{1}{2}(\kappa^{\mu}-\gamma^{\mu}+2l^{\mu})\partial^{4}(\kappa_{\mu}-\gamma_{\mu}+2l_{\mu})\;.
\ee
The remaining part $L_{m}'$ involves only mass terms for the gauge bosons and no mixing
\be\label{DC28}
L_{m}'=C^{\mu\nu\rho}C_{\mu\nu\rho}+v^{(t)\mu}v_{\mu}^{(t)}-\dfrac{3}{2}\tilde{v}\partial^{2}\tilde{v}+\dfrac{4}{9}w^{(t)\mu}w_{\mu}^{(t)}\;.
\ee

The kinetic term for the gauge bosons~\eqref{kinetic Lagrangian for LL gauge field} contributes only for the transversal gauge bosons,
\be\label{DC29}
\!\! L_{F}=-\dfrac{Z}{4}A^{\mu\nu\rho}\partial^{2}P_{\mu}^{\;\,\sigma} A_{\sigma\nu\rho}=-\dfrac{Z}{4}B^{\mu\nu\rho}\partial^{2}B_{\mu\nu\rho}\;.
\ee
The decomposition is similar to the transversal part of eq.~\eqref{DC21}, replacing $m^{2}$ by $-Z\partial^{2}$,
\be\label{DC30}
\begin{split}
L_{F}=\dfrac{Z}{4}\biggl{\lbrace}\dfrac{1}{2}E^{\nu\rho}\partial^{4}E_{\nu\rho}-C^{\mu\nu\rho}\partial^{2}C_{\mu\nu\rho}-v^{(t)\mu}\partial^{2}v_{\mu}^{(t)} \\
+\dfrac{3}{2}\tilde{v}\partial^{4}\tilde{v}-\dfrac{4}{9}w^{(t)\mu}\partial^{2}w_{\mu}^{(t)}+\dfrac{2}{3}\tilde{w}\partial^{4}\tilde{w}\biggr{\rbrace}\;.
\end{split}
\ee

At this point we have expressed both $L_{U}$ and $L_{F}$ in terms of the fields of our decomposition. The transition to momentum space is straightforward. The pieces $\sim q^{4}$ are artefacts of the particular normalization of fields in our decomposition. They can be absorbed by a momentum dependent rescaling which makes all fields in the decomposition of the vierbein dimensionless, and gives all fields in the expansion of the gauge fields the canonical dimension of mass.
The high momentum behavior for all modes is then $\sim q^{2}$, as appropriate for an action that contains only two derivatives. The different blocks can now be diagonalized separately. 

\section{Stability of classical theory\label{sec: Stability of classical theory}}

In this section we diagonalize the propagator. The question of stability of the classical theory can be read off directly from the form of the propagators. Flat space is  stable if there are neither ghosts nor tachyons and all modes are described by massive or massless particles. Stability of flat space extends to stability of the high-momentum modes for arbitrary ``background configurations" of the gauge fields and vierbein.

\subsection{High momentum limit\label{subsec: High momentum limit}}
\smallskip

Before proceeding to the diagonalization of the inverse propagator it is instructive to discuss the limit $q^{2}\rightarrow \infty$. In this limit $L_{U}^{(2)}$ and $L_{U}^{(3)}$ can be neglected. The infinite momentum limit becomes a good approximation for $q^{2}\gg m^{2}$.
In consequence, the leading short distance behavior is well described by an action based on $L_F+L_U^{(1)}$, evaluated for the flat space solution.
This is a free theory with inverse propagators for the physical fluctuations proportional to $q^2$. 
No instability occurs.
For this free theory the analytic continuation to Minkowski space is straightforward.
We therefore consider our pregeometry based on a Yang-Mills theory as a candidate for an ultraviolet completion of quantum gravity. 

Beyond the leading approximation the action \eqref{starting action} also entails interactions.
First, there are the gauge interactions mediated by the gauge field $A_\mu$.
The effective gauge coupling $g$, with $\alpha=g^2/(4\pi)$, is given by
\al{
\alpha =\frac{4\pi}{Z}\,.
}
It vanishes in the limit $Z\to\infty$.
For large $Z$ a perturbative treatment of the gauge interactions becomes possible.

In contrast, the vierbein-mediated ``gravitational interactions" cannot be switched off.
We can use scaling fields~\cite{CWFSI}
\al{
&\tilde e_\mu{}^m=ke_\mu{}^m\,,&
&\tilde g_{\mu\nu}=k^2 g_{\mu\nu}\,,&
&\tilde g^{\mu\nu}=k^{-2}g^{\mu\nu}\,,
}
with $k$ some ``renormalization scale".
As a result one finds
\al{
S=\frac{1}{16\pi} \int_x \tilde e\left( \frac{1}{2\alpha} F_{\mu\nu mn}F_{\rho\sigma}{}^{mn} +\frac{1}{\gamma}\tilde U_{\mu\nu m}\tilde U_{\rho\sigma}{}^m \right) \tilde g^{\mu \rho}\tilde g^{\nu\sigma}\,,
}
with 
\al{
\tilde U_{\mu\nu}{}^m=D_\mu \tilde e_\nu{}^m=k U_{\mu\nu}{}^m\,,
}
and dimensionless coupling
\al{
\gamma =\frac{k^2}{4\pi m^2}\,.
}
A multiplicative rescaling of $\tilde e_\mu{}^m$ does not change the Levi-Civita connection $\Gamma_{\mu\nu}{}^{\sigma}$ in the covariant derivative \eqref{field strength for vierbein}.
It only results in a multiplicative rescaling of $\tilde U_{\mu\nu}{}^m$.
As a result, $\gamma$ is not a free parameter - it can be set to any arbitrary value by a rescaling of $\tilde e_{\mu}{}^m$.
In contrast to the gauge interactions, the gravitational interactions do not switch off in the limit $\gamma\to0$.
Using the freedom of field-rescaling the only remaining free parameter for the classical theory is the gauge coupling $\alpha$.

\subsection{Diagonalization of the propagator\label{subsec: Diagonalization of the propagator}}
\smallskip

Let us next consider the full classical action based on $L_{F}+L_{U}$.
With an appropriate normalization, the inverse propagator for the gauge boson fluctuactions $C$, $v^{(t)}$, $w^{(t)}$ and $\tilde{v}$ is the standard one for massive particles in momentum space
\be\label{DC31}
P=G^{-1}=q^{2}+\dfrac{m^{2}}{Z}\;.
\ee
This is the inverse propagator of a particle with mass $m/\sqrt{Z}$. No instability occurs in this sector. 

For the traceless transversal tensors in the $t-E$ - sector the inverse propagator matrix takes the form
\be\label{DC32}
P=G^{-1}=\dfrac{q^{2}}{4}\begin{pmatrix}
Zq^{2}+m^{2}&, \,m^{2}\\
m^{2}&,\,m^{2}
\end{pmatrix}\;.
\ee
The only zero eigenvalue of $P$ occurs for $q^{2}=0$. The eigenvalues of $P$ are given by
\be\label{DC34}
\lambda_{\pm}=\dfrac{q^{2}}{8}\Big(Zq^{2}+2m^{2}\pm\sqrt{Z^{2}q^{4}+4m^{4}}\,\Bigr)\;.
\ee

These eigenvalues have dimension $\textup{mass}^{4}$. We can switch to the more usual normalization for bosonic fields for which the boson fields have dimension of mass and the inverse propagator has dimension mass squared. This can be achieved by a momentum dependent renormalization of the fields, $E_{R\mu\nu}=\dfrac{q}{2}E_{\mu\nu}$, $q=\sqrt{|q^{2}|}$, $t_{R\mu\nu}=\dfrac{m}{2}t_{\mu\nu}$, resulting in the renormalized inverse propagator matrix
\be\label{DC35}
P_{R}=
\begin{pmatrix}
Zq^{2}+m^{2}&,\;mq\\
mq&,\;q^{2}
\end{pmatrix}\;.
\ee
The $q$- dependence in the renormalization of $E$ corresponds to the standard normalization of gauge fields $A$, noting $A\sim qE$, while the normalization of $t$ provides a canonical mass dimension to this field. 

Poles in the propagator correspond to vanishing eigenvalues of $P_{R}$. The condition for the existence of propagator poles reads
\be\label{TDI Eq: H7}
\det{P_{R}}=Zq^{4}=0\;.
\ee
We conclude that the only possible poles of the propagator occur for $q^{2}=0$.

For the renormalized fields the eigenvalues of~$P_{R}$ for general values of $q^{2}$ are given by
\be\label{DC36}
\begin{split}
&\lambda_{\pm}=\dfrac{1}{2}\Bigl{\lbrace}(Z+1)q^{2}+m^{2}\\ 
&\quad\quad\quad\quad\pm\sqrt{(Z-1)^{2}q^{4}+2(Z+1)q^{2}m^{2}+m^{4} }\Bigr{\rbrace}\;.
\end{split}
\ee
They correspond to the inverse propagators after diagonalization and describe the propagation of independent modes.

For the high momentum behavior, $q^{2}\gg m^{2}$, this yields
\be\label{DC37}
\lambda_{+}=
\begin{cases}
Z\Bigl(q^{2}+\dfrac{m^{2}}{Z-1}\Bigr) \phantom{\Bigg(}&\text{for $Z>1$}\\
q^{2}+\dfrac{m^{2}}{1-Z}&\text{for $Z<1$} \;,
\end{cases}
\ee
and
\be
\lambda_{-}=
\begin{cases}
 q^{2}-\dfrac{m^{2}}{Z-1} \phantom{\Bigg(}&\text{for $Z>1$}\\
 Z\Bigl(q^{2}-\dfrac{m^{2}}{1-Z}\Bigr)&\text{for $Z<1$} \;.
\end{cases}
\ee
The apparent zeros in these approximate expression cannot correspond to poles in the propagators since eq.~\eqref{TDI Eq: H7} tells us that all possible poles occur for $q^{2}=0$.
Near the pole at $q^{2}=0$ one finds
\be\label{DC38}
\lambda_{-}=\dfrac{Zq^{4}}{m^{2}}\quad ,\quad\lambda_{+}=m^{2}+(Z+1)q^{2}\;.
\ee
Again, the apparent zero in $\lambda_{+}$ is a``fake pole", since the only possible poles for $q^{2}=0$ are associated to $\lambda_{-}$. The function $\lambda_{+}(q^{2})$ has no zero. No propagating particle is associated to this mode.

The zero of the inverse propagator at $q^{2}=0$ corresponds to a double pole in the propagator. A propagator $1/q^{4}$ leads to a ``secular instability" after analytic continuation to Minkowski space. It does not pose any problem for a well defined euclidean functional integral. A double pole can be considered as the boundary between stability and instability. The stability for low $q^{2}$ should be discussed within the quantum effective action once quantum fluctuation effects are included. We will see below that additional terms in the effective action modify the behavior for $q^{2}\rightarrow 0$. 

For the other limit at large $q^{2}$ one may assume that the short distance behavior of the effective action is close to the classical action. The short distance fluctuations in the $t$ - $E$ - sector are stable.

We next turn to the remaining part in the sector of physical scalar fluctuations. 
In the $\tilde{w}-\sigma$ - sector the inverse propagator matrix
\be\label{DC39}
P=\dfrac{q^{2}}{3}\begin{pmatrix}
Zq^{2}+m^{2}&,\; -\dfrac{m^{2}}{2}\\
-\dfrac{m^{2}}{2}&\!\! ,\; \phantom{\Bigg|}\phantom{-}\dfrac{m^{2}}{4}\;
\end{pmatrix}
\ee
has the same structure as eq.~\eqref{DC32}, up to an overall~factor~$4/3$ and a rescaling of $\sigma$ by a factor $(-2)$.
The poles in the propagators are therefore the same as in the sector of transverse traceless tensors $(t, E)$. Together with eq.~\eqref{DC31} for the scalar $\tilde{v}$ we observe in the scalar sector one massive particle and one massless particle with a double pole. The scalar $\tilde{u}$ is a gauge degree of freedom of diffeomorphisms and does not appear in the quadratic action. 

The sector of transversal vectors contains several independent blocks. The vectors $v^{(t)}$ and $w^{(t)}$ from the expansion of the transversal gauge field fluctuations correspond to massive particles with propagators given by eq.~\eqref{DC31}. A third physical vector fluctuation is given by the combination
\be\label{TDI Eq. C7}
s_{\mu}=\dfrac{1}{2}(\kappa_{\mu}-\gamma_{\mu})+l_{\mu}\;.
\ee
Its inverse propagator $P=m^{2}q^{4}$ obtains from $L_{\kappa\gamma l} $ in $L_{U}$ as given by eq.~\eqref{DC27}. A canonical normalization of this vector field absorbs in $P$ a factor $m^{2}q^{2}$. We end with a massless vector particle with normalized inverse propagator $P_{R}=q^{2}$. The transverse vector fluctuations contain also two gauge degrees of freedom, one from diffeomorphisms and the other from the $SO(4)$ - gauge symmetry. We may take them as $\kappa_{\mu}$ and $l_{\mu}$ at fixed $s_{\mu}$. Only the physical mode $s_{\mu}$ appears in the quadratic action.

The fluctuations $C$ describe one more massive particle with inverse propagator given by eq.~\eqref{DC31}. What remains is the sector of antisymmetric tensor fluctuations, consisting of the longitudinal gauge bosons $M$ and the antisymmetric vierbein fluctuations $b$. 
For the longitudinal gauge bosons there is no contribution from $L_{F}$. The inverse propagator matrix in the $M-b$ - sector obtains from $L_{bM}$ in eq.~\eqref{DC26}
\be\label{DC40}
P=\dfrac{q^{2}m^{2}}{8}\begin{pmatrix}
\phantom{-}4&-2\\-2&\phantom{-}1
\end{pmatrix}\;.
\ee
After proper renormalization of the fields this becomes
\be\label{DC41}
P_{R}=\begin{pmatrix}
\phantom{\Big|}m^{2}&-mq\\-mq&q^{2}
\end{pmatrix}\;.
\ee
One eigenvalue $\lambda_{+}$ of this matrix corresponds to the physical fluctuation
\be\label{TDI D7}
r_{\mu\nu}=\dfrac{1}{2}b_{\mu\nu}-M_{\mu\nu}\;.
\ee
Only this physical mode appears in $L_{U}$ through $L_{bM}$ in eq.~\eqref{DC26}. It corresponds to a massive particle with inverse propagator
\be\label{DC42}
\lambda_{+}=q^{2}+m^{2}\;.
\ee
The other eigenvalue $\lambda_{-}$ is zero for all $q^{2}$, corresponding to $\det P=0$ or $\det P_{R}=0$. This is the gauge mode of local $SO(4)$ - gauge transformations. Indeed, these gauge transformations, applied to a``vacuum state" $A_{\mu mn}=0$, $e_{\mu}^{m}=\delta_{\mu}^{m}$, do not only shift the longitudinal components of $A_{\mu mn}$, but also rotate $e_{\mu}^{m}$, contributing infinitesimally to~$H_{\mu\nu}^{(A)}$.
For this reason the gauge modes can be taken as linear combinations of $L_{\mu\nu}$ and $b_{\mu\nu}$. We use the freedom in the precise parametrization of the gauge mode to take the fluctuation $L_{\mu\nu}$ at fixed physical fluctuation $s_{\mu\nu}\;$.

In total we observe 9 massless physical modes ($t-E$ , $\tilde{w}-\sigma$ , $ s$), 15 massive physical modes ($v^{(t)},\,w^{(t)},\,\tilde{v},\, C,\, r$) and 10 gauge degrees of freedom ($\kappa,\, u,\, l,\, M$).
We recall that the gauge fluctuations have to be taken at fixed physical modes.
One linear combination in the $t-E$-sector and in the $\tilde{w}-\sigma$-sector (6 modes) correspond to physical modes that do not describe a propagating particle. Similar to Einstein gravity one expects that due to non-local projectors not all of the modes ($t-E\,,\, \tilde{w}-\sigma \,,\, s\,,\, v^{(t)},\;w^{(t)},\;\tilde{v}\,,\, C\,, \,r\,)$ are propagating particles~\cite{CWMF}. This reduces the number of modes corresponding to physical propagating particles below 24. 

\subsection{Short distance completion for euclidean gravity\label{subsec: Short distance}}
\smallskip

Taking things together, for the action $L_{U}+L_{F}$ the propagators of the physical fluctuations around euclidean flat space (excluding the gauge modes) have the following properties: 
\begin{enumerate}[label=(\roman*)]
\item In the complex $q^{2}$- plane all poles occur on the negative real axis, including $q^{2}=0$.\label{Ap.A (i)}
\item For the massive modes with location of the poles at $q_{p}^{2}=-m_{i}^{2}$ the residuum is always positive, as required for standard stable massive particles.\label{Ap.A (ii)}
\item For real $q^{2}\geqslant 0$ the inverse propagator is positive for all modes, increasing $\sim q^{2}$ for $q^{2}\rightarrow\infty$.\label{Ap.A (iii)}
\item At $q^{2}=0$ there are degenerate poles, with eigenvalues of the inverse propagator $\sim q^{4}$. These massless modes are linear combinations of gauge boson - and vierbein-fluctuations. The behavior $\sim q^{4}$ can be considered as the boundary of the region of stability.\label{Ap.A (iv)}
\end{enumerate}
We can consider this model of $SO(4)$ - pregeometry as a valid candidate for the ultraviolet completion of euclidean gravity.

\subsection{Analytic continuation\label{subsec: Analytic continuation}}
\smallskip

For the momentum dependence of the propagators discussed above there is no obstruction for analytic continuation to a Minkowski signature. Investigating the propagator for a Minkowski signature in the complex $q_{0}$ - plane one finds besides the poles branch cuts for $q_{0}^{2}>| q_{c}^{2}|+\vec{q}^{\,2}$, with positive non zero $|q_{c}|^{2}$ to be discussed in more detail in sect.\ref{sec: Graviton Propagator}.
Such a setting is rather promising for a consistent description of quantum gravity.

The general analytic continuation of our gauge theory of pregeometry has to map the compact gauge group $SO(4)$ to the non-compact Lorentz group $SO(1, 3) $. This analytic continuation can be achieved by a complex continuation of the vierbein $e_{\mu}{}^{m}$ and the gauge fields $A_{\mu m n}$ by mapping
\begin{align}
\label{86A}
&e_{\mu}{}^{0}\to e'_{\mu}{}^{0}=e^{i\varphi}e_{\mu}{}^{0}\;,\nn\\
&A_{\mu k 0}\to\A'_{\mu k 0}=e^{-i\varphi}A_{\mu k 0}\;,
\end{align}
while fields carrying only ``Lorentz indices" $k=1,\: 2, \: 3 $ remain invariant, $e'_{\mu}{}^{k}=e_{\mu}{}^{k}$ , $A'_{\mu kl}=A_{\mu kl}$ . For \mbox{$\varphi =\pi/2$} one reaches Minkowski signature of the metric. With 
\mbox{$e'_{\mu}{}^{0}=ie_{\mu}{}^{0}$} , $A'_{\mu m0}=-iA_{\mu m 0}$ , the analytically continued metric reads
\be
\label{86B}
g'_{\mu\nu}=e'_{\mu}{}^{m}e'_{\nu}{}^{n}\delta_{mn}=e_{\mu}{}^{m}e_{\nu}{}^{n}\: \eta_{mn}\;.
\ee
We observe the appearance of the $SO(1, 3)$ invariant tensor $\eta_{mn}=diag(-1, 1, 1, 1)$. Correspondingly, the $SO(4)$ gauge symmetry becomes $SO(1, 3)$ after analytic continuation.

This is seen most easily if we define a Minkowski-vierbein and Minkowski gauge fields by the convention
\be
\label{86C}
e_{ \mu}^{\prime(M)0}=-ie_{\mu}^{\prime\, 0}\;,\quad  A_{\mu\, k0}^{\prime\,(M)}=i A'_{\mu k0}\;.
\ee
The fields $e_{\mu}^{\prime(M)m}$ and $A_{\mu mn}^{\prime\, (M)}$ are real. After analytic continuation with $\varphi=\pi /2$ they equal the original euclidean fields, $e_{\mu}^{\prime(M)m}=e_{\mu}{}^{m}$, $A_{\mu mn}^{\prime\,(M)}=A_{\mu mn}$. The analytic continuation of the metric~\eqref{86B} remains, however, being unaffected by the convention chosen for the vierbein and gauge fields. The analytic continuation of the gauge group follows if one continues the parameters $\varepsilon_{mn}$ of infinitesimal $SO(4)$-transformations by the same rules. More details for the analytic continuation can be found in ref.~\cite{Wetterich:2021hru}.

We emphasize that this procedure of analytic continuation is done at fixed coordinates $x^{\mu}$.
It only concerns the Lorentz-indices $m$ of the fields, not the world indices $\mu$. This version of analytic continuation in the vierbein and metric has been proposed previously~\cite{Wetterich:2010ni}. 
As long as only vierbein and metric are concerned it is equivalent to analytic continuation in the time coordinate. In particular, for the squared momentum, $q^{2}=g^{\mu\nu} q_{\mu}q_{\nu}$, or corresponding expressions with covariant derivatives, the analytic continuation of the metric can be described equivalently with a fixed metric and analytic continuation in the zero component of momentum $q_{0}$. Propagators depending on $q^{2}$ do not change. Analytic continuation in flat space only changes the euclidean expression $q^{2}=q_{0}^{2}+\vec{q}^{\, 2}$.

For theories with gauge fields carrying Lorentz indices the analytic continuation of fields is mandatory, however. If one would only continue the time coordinate, the gauge group would remain $SO(4)$, resulting in a mismatch with the $SO(1, 3)$-symmetry of Minkowski space. The analytic continuation of fields with Lorentz-indices has the additional advantage that it is well defined for an arbitrary geometry. There is a clear rule how the metric transforms, using  $e_{\, \mu}^{\prime(M)m}=e_{\mu}{}^{m}$ and continuing $\delta_{mn}\to\eta_{mn}$, while coordinates remain fixed. Furthermore, fields as $U_{\mu\nu}{}^{\rho}=U_{\mu\nu}{}^{m}e_{m}{}^{\rho}$ remain invariant under analytic continuation since the phase factors cancel out. In short, by use of the``Minkowski convention" for fields  $e_{\, \mu}^{\prime(M)m}$, $A_{\mu mn}^{\prime\, (M)}$, the effects of analytic continuation in the fields can equivalently be described by fixed fields $e_{\mu}{}^{m}$, $A_{\mu mn}$ and analytic continuation in the invariants $\delta_{mn}\to\eta_{mn}$, $\varepsilon_{mnpq}\to i\varepsilon_{mnpq}$ . (The change in the $\varepsilon$-tensor accounts for the appearance of a factor $i$ in the exponent $ e^{-S}$ in the functional integral, due to $e'=ie$ .)

\section{Effective action and emergence of \\general relativity\label{sec: Effective action and emergence of general relativity}}

The action~\eqref{starting action} may be considered as the microphysical or classical action which is used to define a functional integral. Quantum fluctuations will lead to a macroscopic or effective action. The quantum effective action is the generating functional of the one-particle-irreducible Green's functions. It includes all fluctuation effects. The field equations obtained by variation of the effective action are exact. They replace the field equations obtained from the classical action. The second functional derivative of the effective action is the inverse propagator in an arbitrary background of fields.

With a proper definition (and in the absence of anomalies) the effective action has the same symmetries as the classical action. In general, this invariant functional of the fields contains infinitely many invariants. The arguments are now macroscopic fields, corresponding to expectation values of microscopic fields in the presence of arbitrary sources.

We do not attempt to compute the effective action $\Gamma$ in this paper. We rather make an ansatz for its form which is consistent with the symmetries. In practice, we replace $S$ by $\Gamma$ and add further invariants. The parameters $m^{2}$ and $Z$ are replaced by renormalized parameters which include the effects of the quantum fluctuations. For the additional invariants we investigate a derivative expansion up to second order in the derivatives. It may be possible to include some of the additional invariants into the classical action. We will leave it open here if they are also part of the classical action or if they are generated only by fluctuation effects.

\subsection{New invariants\label{subsec: New invariants}}
\smallskip

At low momenta a derivative expansion of the effective action is often a good guide. We will next add to the action~\eqref{starting action} further invariants with up to two derivatives, 
\be\label{EA1}
\Gamma =\int d^{4} x e ( L_{F}+L_{U}+U+L_{R}+L_{G})\;.
\ee
The term without derivatives is a cosmological constant $U$. We will first neglect it here for simplicity, such that flat space remains a solution of the field equations. This may be motivated by the observation that in the present universe flat space is a very good approximation. For early cosmology this may no longer hold. We discuss the effects of $U\neq 0$ in sect.~\ref{sec: Cosmological constant}.

The particular field content with a vierbein permits an invariant that is linear in the fields 
\be\label{RR bisher 35}
L_{R}=-\dfrac{M^{2}}{2}F_{\mu\nu}{}^{\mu\nu}\;.
\ee
Indeed, contraction of the field strength with the inverse vierbein yields a scalar which is invariant under \mbox{$SO(4)$ - gauge} transformations,
\be\label{TDI K71}
F=F_{\mu\nu}{}^{\mu\nu}=F_{\mu\nu mn} e^{m\mu} e^{n\nu}\;.
\ee
The presence of $L_{R}$ in the effective action is a crucial ingredient for the emergence of general relativity.

Further invariants with two derivatives can be constructed by use of the contractions $F$ and
\be\label{TDI K72}
F_{\mu\nu}=F_{\mu\rho\nu}{}^{\rho}=F_{\mu\rho mn} e_{\nu}{}^{m} \: e^{n\rho}\;,
\ee
as well as
\be 
\label{90A} 
U_{\mu}{}^{\mu}{}_{\rho}=g^{\mu\nu} U_{\mu\nu\rho}\; ,
\ee
namely
\be\label{TDI K73}
L_{G}=\dfrac{B}{2}F_{\mu\nu}F^{\mu\nu}+\dfrac{C}{2}F^{2}+\dfrac{n^{2}}{2}U_{\mu}{}^{\mu}{}_{\rho} \: U_{\nu}{}^{\nu\rho}  \;.
\ee
The invariants $\sim C$ and $\sim n\2$ will play a role for the stability of scalar excitations discussed in sect.~\ref{sec: Stability of flat space}. To keep formulas short we set $B=0$ . There exist further invariants with two derivatives partly violating parity or time reversal. They are omitted here. The ``observable" parameters of the model are the dimensionsless couplings or ratios as $Z$, $B$, $C$, $M^{2}/m^{2}$ and $n^{2}/m^{2}$.

\subsection{Field equations\label{subsec: Field equations}}
\smallskip

The field equations obtain by variation of the effective action~\eqref{EA1} with respect to $A_{\mu mn}$ and $e_{\mu}{}^{m}$. For a given effective action $\Gamma$ they are exact, including all effects of quantum fluctuations. Approximations occur only because the effective action is not known precisely. For a model that is thought to describe quantum gravity these field equations have to lead, after analytic continuation to Minkowski signature, to the gravitational field equations that we use on Earth or for the description of black holes.

The field equation for the gauge fields is given by
\be\label{F1}
ZD_{\nu}F^{\mu\nu mn}+2C\partial_{\nu}F(e^{m\mu}e^{n\nu}-e^{m\nu}e^{n\mu})=J^{\mu mn}\;,
\ee
with ``source" or ``current"
\ba\label{F2}
&J^{\mu mn}=m^{2}U^{\mu mn}\nn\\  
&\quad +(2CF-M\2)       (U^{m\mu n}- U^{n\mu m} +U_{\rho}{}^{\rho m}e^{n\mu}-U_{\rho}{}^{\rho n} e^{m\mu}) \nn\\ 
 &\quad -n\2(U_{\rho}{}^{\rho m}e^{n\mu}-U_{\rho}{}^{\rho n} e^{m\mu})  \;.
\end{align}
From the identity $D_{\mu}D_{\nu}F^{\mu\nu mn}=0$ one infers  for \mbox{$C=0$} that solutions of the field equations need a covariantly conserved current,
\be\label{F3}
D_{\mu}J^{\mu mn}=0\;.
\ee

The field equation for the vierbein can be split into symmetric and antisymmetric parts. The one for the symmetric part reads
\be\label{F4}
\dfrac{M^{2}}{2}(F_{\mu\nu}+F_{\nu\mu}-Fg_{\mu\nu})=T_{\mu\nu}^{(U)}+T_{\mu\nu}^{(F)}\;,
\ee
where the r.h.s. involves the symmetric tensors
\ba\label{F5}
&\!\! T_{\mu\nu}^{(U)}=\dfrac{m^{2}}{2}(D_{\rho}U_{\mu\nu}{}^{\rho}+D_{\rho}U_{\nu\mu}{}^{\rho}+U_{\mu}{}^{\tau\rho}U_{\nu\tau\rho}\\
&-\dfrac{1}{2}U^{\sigma\tau\rho}U_{\sigma\tau\rho}g_{\mu\nu})+\dfrac{n\2}{2}\Big{[} 2D_{\rho}U_{\tau}{}^{\tau\rho}g_{\mu\nu}-D_{\mu}U_{\rho}{}^{\rho}{}_{\nu} \nn\\
&-D_{ \nu}U_{\rho}{}^{\rho}{}_{ \mu}+(U_{\mu\nu\rho}+U_{\nu\mu\rho})U_{\tau}{}^{\tau\rho}-U_{\rho}{}^{ \rho}{}_{\sigma} U_{\tau}{}^{\tau\sigma}g_{\mu\nu}\Big{]}\nn
\;,
\end{align}
and
\be\label{F6}
\begin{split}
T_{\mu\nu}^{(F)}=&\dfrac{Z}{2}(F_{\mu}{}^{\rho mn}F_{\nu\rho mn}-\dfrac{1}{4}F^{\sigma\rho mn} F_{\sigma\rho mn} g_{\mu\nu}) \\
&+C\Big{[} F(F_{\mu \nu}+F_{\nu\mu})-\dfrac{1}{2} F\2 g_{\mu\nu}\Big{]}      \;.
\end{split}
\ee
We observe that $T_{\mu\nu}^{(F)}$ is traceless
\be\label{F6A}
g^{\mu\nu}T_{\mu\nu}^{(F)}=0\;.
\ee
The antisymmetric part yields
\ba\label{F7}
m^{2}&D_{\rho}U^{\rho}{}_{\mu\nu}+(M^{2}-2CF)(F_{\mu\nu}-F_{\nu\mu})\\
&-n\2\Big{[}(U_{\mu\nu\rho}-U_{\nu\mu\rho})U_{\tau}{}^{\tau\rho}-D_{\mu}U_{\rho}{}^{\rho}{}_{\nu}+D_{\nu}U_{\rho}{}^{\rho}{}_{\mu}\Big{]}=0\;.\nn
\end{align}
We emphasize that these field equations hold both for euclidean and Minkowski signature. 

\subsection{Simple solutions\label{subsec: Simple solutions}}
\smallskip

An interesting class of solutions is characterized by
\be\label{F8}
U_{\mu\nu\rho}=0\;,\quad\quad D_{\mu}e_{\nu}{}^{m}=0\;.
\ee
In this case one has $F_{\mu\nu\rho\sigma}=R_{\mu\nu\rho\sigma}$, with $R_{\mu\nu\rho\sigma}$ the curvature tensor constructed from the metric $g_{\mu\nu}$.
The field equations~\eqref{F1} and~\eqref{F4} become
\begin{align}\label{F9}
&ZD_{\nu}R^{\mu\nu\rho\sigma}+2C(\partial^{\sigma}Fg^{\mu\rho} - \partial^{\rho}Fg^{\mu\sigma})         =0\;, \nn\\
&M^{2}(R_{\mu\nu}-\dfrac{1}{2}Rg_{\mu\nu})=T_{\mu\nu}^{(F)}\;,\nn\\
&T_{\mu\nu}^{(F)}=\dfrac{Z}{2}(R_{\mu}{}^{\rho\sigma\tau}R_{\nu\rho\sigma\tau}-\dfrac{1}{4}R^{\sigma\rho\tau\eta}R_{\sigma\tau\rho\eta}g_{\mu\nu}) \nn\\
&\quad\quad\quad +CR(2R_{\mu\nu}-\dfrac{1}{2}Rg_{\mu\nu})         \;,
\end{align}
with $R_{\mu\nu}=R_{\mu\rho\nu}{}^{\rho}$, $R=R_{\mu}^{\mu} \ $. Eq.~\eqref{F7} is obeyed identically. 
The second equation~\eqref{F9} is Einstein's equation with an effective energy momentum tensor reflecting the effects of higher curvature invariants.
Taking the trace,
\be\label{F10}
R=0\;,
\ee
the Bianchi identity implies
\be\label{F11}
D^{\nu}T_{\mu\nu}^{(F)}=0\;,
\ee
which reads for $R=0$
\be\label{F12}
\partial_{\mu}(R^{\sigma\tau\rho\eta}R_{\sigma\tau\rho\eta})=4(D^{\nu}R_{\mu}{}^{\sigma\tau\rho})R_{\nu\sigma\tau\rho}+4R_{\mu}{}^{\sigma\tau\rho}D^{\nu}R_{\nu\sigma\tau\rho}
\;.
\ee
A particular solution is flat space, $R_{\mu\nu\rho\sigma}=0$.

For momenta and curvature much smaller than $m^{2}$ eq.~\eqref{F8} becomes a valid approximation, with corrections suppressed by factors $R/m^{2}$ or $D^{2}/m^{2}$. (We assume implicitly that $n\2$ is not much larger in size than $m\2$.) Furthermore, $T_{\mu\nu}^{(F)}$ can be neglected. One ends with Einstein's equation in vacuum. The coupling to elementary particles, that we do not discuss in the present paper, adds an energy momentum tensor as for general relativity.

\subsection{Emergent general relativity\label{subsec: Emergent general relativity}}
\smallskip

For a discussion of the low energy effective theory the invariant $L_{R}$ will become dominant. For $U=0$ this is the leading term in an expansion in derivatives. We will next construct an effective action that is valid for momenta and curvature much smaller than $m\2$. This will reveal the emergence of general relativity with the Einstein-Hilbert action dominating the long distance physics.

For the construction of an effective theory at low momenta $q^2\ll m^2$ we write 
\al{
A_{\mu\nu\rho}= \omega_{\mu\nu\rho} -U_{\mu\nu\rho}\,,
}
and solve the field equation for the tensor $U_{\mu\nu\rho}$ as a functional of the vierbein $e_\mu{}^m$ which is kept fixed.
If the invariant $L_U$ dominates, the solution is simply $U_{\mu\nu\rho}=0$.
The vierbein is then covariantly conserved, $D_\mu e_\nu{}^m=0$, and the gauge fields cease to be independent degrees of freedom, given by $A_{\mu mn}=\omega_{\mu mn}$ . This does not change if we add the term $\sim n\2$ from $L_{G}$.

For the field strength one has~\citep{CWGG}
\al{
F_{\mu\nu\rho\sigma}= R_{\mu\nu\rho\sigma} -V_{\mu\nu\rho\sigma}\,.
}
The curvature tensor involves two derivatives of the vierbein. 
The square of the tensor
\al{
V_{\mu\nu\rho}{}^\sigma =&e_m{}^\sigma \left( D_\mu U_{\nu\rho}{}^m -D_\nu U_{\mu \rho}{}^m \right)  \\ =&D_{\mu}U_{\nu\rho}{}^{\sigma}-D_{\nu}U_{\mu\rho}{}^{\sigma}-U_{\mu}{}^{\sigma\tau}U_{\nu\rho\tau}+U_{\nu}{}^{\sigma\tau}U_{\mu\rho\tau}\nn
}
provides for a kinetic term for $U_{\mu\nu\rho}$ through $L_F$ and $L_{G}$. 
The invariants $L_{F}+L_{G}$ also involve terms linear in $V_{\mu\nu\rho}{}^{\sigma}$ and therefore linear in $U_{\mu\nu}{}^{m}\,$.
For the demonstration of emergent general relativity we omit for a moment the invariant $L_{G}$, cf. ref.~\cite{CWPC} for a more complete discussion.

In the presence of $L_F$ the term linear in $U_{\mu\nu}{}^{m}$ implies that $U_{\mu\nu\rho}=0$ is no longer an exact solution.
This linear term arises from the second term in the expression
\al{
\!\!\!\!\! L_F =\frac{Z}{8}\bigg\{\! R_{\mu\nu\rho\sigma}R^{\mu\nu\rho\sigma} \!-2R_{\mu\nu\rho\sigma} V^{\mu\nu\rho\sigma} \!+V_{\mu\nu\rho\sigma} V^{\mu\nu\rho\sigma} \!\bigg\}\!.\!\!\!\!
}
After partial integration we can replace
\al{
2R_{\mu\nu\rho\sigma} V^{\mu\nu\rho\sigma}\rightarrow -4 U_{\nu\rho}{}^m D_\mu R^{\mu\nu\rho}{}_m\,,
\label{TDI Eq: 40 in F}}
such that the term linear $U_{\mu\nu\rho}$ involves covariant derivatives of the curvature tensor. Similarly, a linear term arises from $L_R$,
\al{
L_R=-\frac{M^2}{2}\left( R -V_{\mu\nu}{}^{\mu\nu} \right)\;,
}
with $R$ the curvature scalar. 

The low momentum effective theory becomes valid if the derivatives of the curvature tensor are small as compared to $m^3$.
In this case the quadratic term for $U_{\mu\nu\rho}$ is dominated by $L_U$ and one finds for small $M^{2}/m^{2}$ the approximate solution
\al{
U_{\mu\nu\rho}=-\frac{Z}{m^2} D^\sigma R_{\sigma \mu \nu\rho}\,.
}
Insertion into $L_F$ and $L_U$ yields a term $\sim (Z^2/m^2)(DRDR)$ with six derivatives that we may neglect for the low momentum effective theory. For $M^{2}$ of the same order as $m^{2}$ the result changes quantitatively, but not qualitatively.
As a result, we can use $U_{\mu\nu\rho}=0$ and find the effective action for low momenta
\al{
S=\int_x \,e \left\{ -\frac{M^2}{2}R +\frac{Z}{8} R_{\mu\nu\rho\sigma}R^{\mu\nu\rho\sigma}+\dfrac{C}{2}R\2 \right\}\,.
\label{TDI Eq: 43}}
Here we have restituted the contribution from $L_{G}$ .
With $e=\sqrt{g}$, $g=\det(g_{\mu\nu})$, the low energy effective action only involves the metric.
General relativity emerges in this limit.

Indeed, the first term in eq.~\eqref{TDI Eq: 43} constitutes the Einstein-Hilbert action and we can associate $M$ with the (reduced) Planck mass. The second term involves the squared Riemann tensor and the squared curvature scalar. Since it contains four derivatives of the metric its influence at momenta $q^{2}\ll M^{2}/Z$ will be suppressed. We conclude that for most practical purposes, except perhaps for very early cosmology or the vicinity of the singularity in black holes, our model predicts precisely the field equations for general relativity. Since general relativity is compatible with all present tests of the gravitational interactions, the same holds for our model of pregeometry. 

\subsection{Stability for low momenta\label{subsec: Stability for low momenta}}
\smallskip

The issue of stability arises on two different levels. The first concerns the properties of the classical action. This topic is related to the question if the euclidean functional integral is well defined. It typically concerns the microscopic or high momentum physics. We have discussed this issue in sect.~\ref{sec: Stability of classical theory}. The second topic concerns the stability of solutions of the field equations. Since the relevant field equations obtain by variation of the effective action this part of the stability discussion concerns the effective action. In particular, for the stability at momenta below the Planck mass the term $L_{R}$ will play an important role. We therefore extend the stability discussion of sect.~\ref{sec: Stability of classical theory} to an effective action based on $L_{R}+L_{U}+L_{F}+L_{G}$.

In contrast to $L_{F}$ and $L_{U}$ the invariant $L_{R}$ is not a square and can therefore take negative values. Since $L_{R}$ is linear in $q$, it cannot modify the leading behavior $\sim{q}^{2}$ for large $q^{2}$. In the infrared limit of small $q^{2}$ it can play an important role, however. One expects that it dominates the behavior near massless propagator poles at $q^{2}=0$.
We will also consider the possibility that $C$ or $n\2$ in $L_{G}$ take negative values. As a part of the classical action this would destroy the boundedness if $\tilde{m}\2 $ or $\tilde{Z}$ are negative,
\be
\label{112A}
\tilde{m}^{2}=m\2+3n\2 \;,\quad\tilde{Z}=Z+12C\;.
\ee
All couplings in the effective action become running couplings. While for the short distance effective action for large $q\2$ one would suggest $\tilde{m}\2 >0$, $\tilde{Z}>0$ in order to make a simple association with the classical action, the stability discussion concerns typically the range $q\2\approx M\2$, or $q\2\ll M\2$. In this range $\tilde{m}\2$ and $\tilde{Z}$ can be negative. We find that such negative values are required for stability in the scalar sector.

For a stability analysis we consider here small fluctuations around flat space. As an important requirement for a realistic theory there should be no tachyonic modes. Otherwise small deviations from flat space would grow exponentially, in contradiction with the observation that cosmology has approached at present flat space with high accuracy. This requirement is based on analytic continuation and realistic cosmology. For euclidean signature a tachyonic mode implies metastability of flat space since the euclidean action for some of the small fluctuations is smaller than the euclidean action for flat space.

\subsection{Expansion of $L_{R}$ on flat space \label{subsec:Expansion of LR on flat space}}
\smallskip

The term $L_{R}$ dominates the small momentum behavior. Its contribution is therefore a central ingredient for the analysis of stability for the effective action.
We need to expand
\be\label{DC43}
eL_{R}=-\dfrac{M^{2}}{2}ee^{m\mu}e^{n\nu}F_{\mu\nu m n}
\ee
in quadratic order in $H$ and $A$ around $e_{\mu}^{m}=\delta_{\mu}^{m}\,,\, A_{\mu mn}=0$.
Since $F_{\mu\nu mn}$ is a least linear in $A_{\mu mn}$ according to eq.~\eqref{TDI2}, we need the vierbein in linear order in $H$,
\ba\label{DC44}
&ee^{m\mu}e^{n\nu}=\delta^{m\mu}\delta^{n\nu}\\
&\quad\quad \quad \quad  +\dfrac{1}{2}H_{\rho\sigma}\bigl(\delta^{m\mu}\delta^{n\nu}\delta^{\rho\sigma}-\delta^{m\mu}\delta^{n\rho}\delta^{\nu\sigma}-\delta^{m\rho}\delta^{n\nu}\delta^{\mu\sigma}\bigr)\;. \nn
\end{align}
The linear term $\partial_{\mu}A_{\nu}^{\quad\mu\nu}-\partial_{\nu}A_{\mu}^{\quad\mu\nu}$ is a total derivative and therefore vanishes.

In quadratic order one finds two contributions, \mbox{$L_{R}=L_{R}^{(1)}+L_{R}^{(2)}$,} 
\ba\label{DC45}
&eL_{R}^{(1)}=-\dfrac{M^{2}}{4}H_{\rho\sigma}\Bigl{\lbrace}\bigl(\partial_{\mu}A_{\nu}^{\quad\mu\nu}-\partial_{\nu}A_{\mu}^{\!\!\quad\mu\nu}\bigr)\delta^{\rho\sigma}\\
&\quad\quad -\bigl(\partial_{\mu}A_{\nu}^{\!\!\quad\mu\rho}\!\!-\partial_{\nu}A_{\mu}^{\!\!\quad\mu\rho}\bigr)\delta^{\nu\sigma}\!\!-\!\bigl(\partial_{\mu}A_{\nu}^{\!\!\quad\rho\nu}-\partial_{\nu}A_{\mu}^{\!\!\quad\rho\nu}\bigr)\delta^{\mu\sigma}\!\Bigr{\rbrace}\;,\nn
\end{align}
and
\be\label{DC46}
eL_{R}^{(2)}=-\dfrac{M^{2}}{2}\Bigl{\lbrace}A_{\mu}^{\quad\mu\rho}A_{\nu\rho}^{\quad\nu}-A_{\nu}^{\quad\mu\rho}A_{\mu\rho}^{\quad\nu}\Bigr{\rbrace}\;.
\ee
The first term involves the term linear in $A_{\mu}$ in $F$ and the linear term~\eqref{DC44}, while the second corresponds to the term quadratic in $A_{\mu}$ in $F$. 

For the first term we observe that only the transversal gauge bosons contribute to $F_{\mu\nu mn}$ in linear order,
\begin{align}\label{DC47}
\!\!\!\!eL_{R}^{(1)}\!=&\dfrac{M^{2}}{2}H_{\rho\sigma}\Bigl{\lbrace}\delta^{\rho\sigma}\partial_{\mu}B_{\nu}^{\!\!\quad\nu\mu}-\partial_{\mu}B^{\sigma\rho\mu}\!-\partial^{\sigma}B_{\mu}^{\!\!\quad\mu\rho}\Big{\rbrace}\phantom{\biggr{\rbrace}}\!\!\!\!\!\!\!\! \nn\\
=&\dfrac{M^{2}}{2}H_{\rho\sigma}\biggl{\lbrace}\Bigl(\dfrac{2}{3}\delta^{\rho\sigma}+\dfrac{1}{3}\dfrac{\partial^{\rho}\partial^{\sigma}}{\partial^{2}}\Bigr)\tilde{b}-\tilde{B}^{(S)\sigma\rho}-\tilde{B}^{(A)\sigma\rho}\nn\\ 
&-\dfrac{2}{3}\partial^{\sigma}w^{(t)\rho}-\partial^{\sigma}\partial^{\rho}\tilde{w}\biggr{\rbrace}\;.
\end{align}
The decomposition yields
\ba\label{DC48}
eL_{R}^{(1)}=&\dfrac{M^{2}}{2}\biggl{\lbrace}\dfrac{2}{3}\tilde{w}\partial^{2}\sigma+\dfrac{1}{2}E^{\rho\sigma}\partial^{2}t_{\sigma\rho}\\
&+\dfrac{1}{2}\varepsilon^{\mu\nu\rho\sigma}v_{\mu}^{(t)}\partial_{\nu}b_{\rho\sigma}
+\dfrac{2}{3}w^{(t)\mu}\partial^{2}(\kappa_{\mu}-\gamma_{\mu})\biggr{\rbrace}\;. \nn
\end{align}
This contributes to the mixing between transversal gauge bosons and vierbein fluctuations, similar to eq.~\eqref{DC20}. As it should be, the gauge modes $\kappa_{\mu}+\gamma_{\mu}$ and $u$ do not appear. 

The second term decomposes as
\ba\label{DC49}
&eL_{R}^{(2)}=\dfrac{M^{2}}{2}\biggl{\lbrace}\dfrac{1}{4}E^{\mu\nu}\partial^{2}E_{\mu\nu}+C_{\mu\nu\rho}C^{\nu\rho\mu}\nn \\
&\quad\quad\quad+\dfrac{2}{9}w^{(t)\mu}w_{\mu}^{(t)}-\dfrac{2}{3}\tilde{w}\partial^{2}\tilde{w}
+\dfrac{1}{2}v^{(t)\mu}v_{\mu}^{(t)}-\dfrac{3}{2}\tilde{v}\partial^{2}\tilde{v}\nn \\
&\quad\quad\quad+\dfrac{4}{3}w^{(t)\mu}\partial^{2}l_{\mu}-\varepsilon^{\mu\nu\rho\sigma}v_{\mu}^{(t)}\partial_{\nu}M_{\rho\sigma}\biggr{\rbrace}\;. 
\end{align}
The different contributions of $eL_{R}$ can be listed as follows
\be\label{DC50}
eL_{R}=\dfrac{M^{2}}{4}\bigl(\Delta L_{tE}+\Delta L_{\sigma\tilde{w}}+L_{vM}+L_{wL}+L_{C}+L_{\tilde{v}}\bigr)\;.
\ee

In the sector of transversal traceless tensors we find a further contribution to the $t$ - $E$ mixing and the term quadratic in $E$, 
\be\label{DC51}
\Delta L_{tE}=E^{\mu\nu}\partial^{2}t_{\mu\nu}+\dfrac{1}{2}E^{\mu\nu}\partial^{2}E_{\mu\nu}\;.
\ee
It has the opposite sign as compared to the contributions from eq.~\eqref{DC24}. In the scalar sector a term
\be\label{DC52}
\Delta L_{\sigma\tilde{w}}=\dfrac{4}{3}(\tilde{w}\partial^{2}\sigma-\tilde{w}\partial^{2}\tilde{w})\;,
\ee
adds to eqs.\eqref{DC24}\eqref{DC25} further contributions.
The term
\be\label{DC53}
L_{vM}=\varepsilon^{\mu\nu\rho\sigma}v_{\mu}^{(t)}\partial_{\nu}(b_{\rho\sigma}-2M_{\rho\sigma})+v^{(t)\mu}v_{\mu}^{(t)}
\ee
can be combined with $L_{bM}$ in eq.~\eqref{DC26}, involving the same physical combination $b_{\mu\nu}-2M_{\mu\nu}=2r_{\mu\nu}$. It induces a new momentum-dependent mixing between $r_{\mu\nu}$ and $v_{\mu}^{(t)}$.
Similarly
\be\label{DC54}
L_{wl}=\dfrac{4}{3}w^{(t)\mu}\partial^{2}(\kappa_{\mu}-\gamma_{\mu}+2l_{\mu})+\dfrac{4}{9}w^{(t)\mu}w_{\mu}^{(t)\mu}
\ee
combines with $L_{\kappa\gamma l}$ in eq.~\eqref{DC27}. It mixes the physical transversal vector fluctuations $s_{\mu}$ and $w_{\mu}^{(t)}$. The remaining parts,
\be\label{DC55}
L_{C}=2C_{\mu\nu\rho}C^{\nu\rho\mu}\;,\quad L_{\tilde{v}}=-3\tilde{v}\partial^{2}\tilde{v}\;,
\ee
add mass terms to the gauge boson fluctuations $C$ and $\tilde{v}$.

\subsection{Expansion of $L_{G}$ on flat space\label{subsec:Expansion of LG on flat space}}
\smallskip

Contributions from $L_{G}$ will be essential in order to avoid tachyons in the scalar sector. The expansion of $eL_{G}$ reads
\be 
\label{125A}
\begin{split}
eL_{G}=&2C\tilde{w}^{*}\partial^{4}\tilde{w}-\dfrac{n\2}{8}(\sigma^{*}-2w^{*})\partial\2 (\sigma-2w)\\
&+\dfrac{n\2}{8}\gamma_{\mu}^{*}\partial^{4}\gamma^{\mu}+\dfrac{2n\2}{9}w_{\mu}^{(t)*}w^{(t)\mu}\\
&-\dfrac{n\2}{6}\Big{(}w_{\; \mu}^{(t)*}\partial\2 \gamma^{\mu}+\gamma_{\mu}^{*}\partial\2 w^{(t)\mu}\Big{)}\;.
\end{split}
\ee
Thus $L_{G}$ does not contribute to the transverse traceless tensor modes. It yields additional terms in the scalar and vector sector, contributing to the blocks $L_{\sigma\tilde{w}
}$ and $L_ {wl}$.

Adding the contributions of $L_{R}+L_{G}$ to the ones from $L_{U}+L_{F}$ the stability discussion can again be performed separately for different blocks. We need to compute the propagators from a diagonalization of the individual blocks. Stability depends on the behavior of these propagators. In the next section we will compute the propagator in the transverse traceless tensor sector which describes the massless graviton and a massive spin-two particle. In the following sect.~\ref{sec: Stability of flat space} we turn to the scalar sector. There we also discuss briefly the other sectors.

\section{Graviton propagator\label{sec: Graviton Propagator}}

We are now in a position to discuss the propagator of the graviton. In the low momentum limit we expect an inverse propagator $\sim q^{2}$ according to the emergence of general relativity discussed in the preceding section. Also for high $q^{2}$ one has a propagator $\sim q^{2}$ according to the findings of sect.~\ref{sec: Stability of classical theory}. This momentum region is not affected by the addition of $L_{R}$, which is a subleading correction for $q^{2}\gg M^{2}$. One expects some smooth interpolation between the two limits. In the view of later analytic continuation we discuss the graviton propagator for arbitrarily complex $q^{2}$.

\subsection{Inverse propagator for transversal traceless\\ tensors\label{subsec: Inverse propagator for transversal traceless tensors}}
\smallskip

Our model has two fields transforming as transverse traceless tensors, namely $t_{\mu\nu}$ and $E_{\mu\nu}$. They mix, and the graviton has to be identified with one of the eigenmodes of the inverse propagator matrix. In the presence of $L_{R}$ the renormalized inverse propagator matrix~\eqref{DC35} in the \mbox{$t-E$ - sector} is replaced by
\be\label{DC56}
P_{R}^{(tE)}=\begin{pmatrix}
Zq^{2}+m^{2}-M^{2}&,&\;\dfrac{m^{2}-M^{2}}{m}q\phantom{\Bigg|}\\
\dfrac{m^{2}-M^{2}}{m}q & , & q^{2}\phantom{\Bigg|}
\end{pmatrix}\;.
\ee
This matrix is the central quantity for the discussion of the graviton propagator. In addition to the massless graviton it will describe a massive particle with mass $\sim M$. For $M\rightarrow 0$ both excitations are massless, as found in sect.~\ref{sec: Stability of classical theory}.

The poles of the propagators or zero eigenvalues of~$P_{R}$ occur for
\be\label{DC57}
q^{2}=0\;,\quad q^{2}=-\mu^{2}\;,
\ee
with
\be\label{DC58}
\mu^{2}=\dfrac{m^{2}}{Z}y(1-y)\;,\quad y=\dfrac{M^{2}}{m^{2}}\;.
\ee
This is seen easily from the condition $\det (P_{R})=0$. No other poles of the propagator are possible. For $M^{2}\rightarrow 0$ the second pole moves to zero, 
\be\label{TDI Eq H8}
\mu^{2}=\dfrac{M^{2}}{Z}\Bigl(1-\dfrac{M^{2}}{m^{2}}\Bigr)\;.
\ee
A stable particle requires $\mu^{2} \geqslant 0$, or
\be\label{DC59}
0\leqslant y \leqslant 1\;,\quad 0\leqslant M^{2}\leqslant m^{2}\;.
\ee
Otherwise one encounters a tachyonic instability. The constraint~\eqref{DC59} is a first condition for stable gravity.

The inverse particle propagators are given by the eigenvalues of~$P_{R}$,
\ba\label{DC60}
&\lambda_{\pm} = \dfrac{1}{2}\Biggl{\lbrace} (Z+1)q^{2}+m^{2}-M^{2} \\
&\quad\quad\;\pm\sqrt{\bigl[ (Z-1)q^{2}+m^{2}-M^{2}\bigr]^{2}+4\dfrac{q^{2}}{m^{2}}(m^{2}-M^{2})^{2}}\;\Biggr{\rbrace}\;.\nn
\end{align}
All properties of the propagation of transverse traceless tensor modes in flat space are encoded in these two eigenvalues. For suitable ranges of parameters we will find that the eigenvector to $\lambda_{-}$ describes the massless graviton, while the eigenvector to $\lambda_{+}$ accounts for a stable massive particle. At long distances only the graviton will matter. 

\subsection{Graviton\label{subsec: Graviton}}
\smallskip

Indeed, for $q^{2}\rightarrow 0$ the eigenvalue $\lambda_{-}(q^{2})$ is the inverse propagator for the massless graviton, corresponding to a Taylor expansion
\be\label{DC61}
\lambda_{-}=\dfrac{M^{2}q^{2}}{m^{2}}+\bigl(Z-\dfrac{M^{2}}{m^{2}}\bigr)\dfrac{q^{4}}{m^{2}}+\dots
\ee
A factor $4/m^{2}$ may be absorbed by a normalization which makes the graviton fluctuation dimensionless. The relation between the expansion~\eqref{DC61} and the low-momentum effective action~\eqref{TDI Eq: 43} involves a wave function renormalization. We can write
\be\label{139A}
\lambda_{-}=\frac{M^{2}q^{2}+Zq^{4}+\dots}{m^{2}+q^{2}}
\ee
and absorb the factor $(m^{2}+q^{2})^{-1}$ into the normalization. Indeed, the tensor fluctuation $t_{\mu\nu}$ in the effective low-momentum theory equals the eigenfunction of $\lambda_{-}$ only in lowest order in an expansion in $q^{2}$. For the effective low-momentum theory one employs $U_{\mu\nu\rho}=0$ or $E_{\mu\nu}=-t_{\mu\nu}$, while the eigenfunction of $\lambda_{-}$ obeys this relation only in lowest order in $q^{2}$. The use of the low-energy effective action away from solutions of the field equations (``off shell") has limitations.

For the other eigenvalue one has
\be\label{TDI Eq: C9}
\lambda_{+}(q^{2}=0)=m^{2}-M^{2}\;.
\ee
For $M^{2}>0$ and $M^{2}<m^{2}$ only a single pole remains at $q^{2}=0$. The degeneracy of the double pole in eq.~\eqref{DC38} is lifted. For $M^{2}=m^{2}$ one finds two single massless poles at $q^{2}=0$.

Near the second pole at $q_{p}^{2}=-\mu^{2}$ one finds for
\be\label{DC62}
-(Z+1)\mu^{2}+m^{2}-M^{2}<0\;,
\ee
or
\be\label{DC63}
Z<Z_{c}\;,\quad Z_{c}=\dfrac{y}{1-y}\;,
\ee
a negative $\lambda_{-} $ for $q^{2}$ near $-\mu^{2}$,
\be\label{DC64}
\lambda_{-}=-m_{t}^{2}-\dfrac{1}{2}\Bigl(Z+1-\dfrac{A_{t}m^{2}}{m_{t}^{2}}\Bigr)(q^{2}+\mu^{2})\;.
\ee
Indeed, for $Z<Z_{c}$, the combination $m_{t}^{2}$ is positive, 
\be\label{DC65}
m_{t}^{2}=(Z+1)\mu^{2}-m^{2}+M^{2}=m^{2} y(1-y)\Bigl(\dfrac{1}{Z}-\dfrac{1}{Z_{c}}\Bigr)\;.
\ee
The coefficient $A_{t}$ is given by
\be\label{DC66}
A_{t}=(1-y)\biggl(Z+1-\Bigl(Z+\dfrac{1}{Z}\Bigr) y\biggr)\;.
\ee
Here we have always assumed the range $0\leqslant y\leqslant 1$ required by stability according to eq.~\eqref{DC59}. With the conditions~\eqref{DC59}~and~\eqref{DC63} the eigenvalue $\lambda_{-}(q^{2})$ therefore has a single zero, located at $q^{2}=0$. There is no tachyonic or ghost pole in the graviton propagator.

For large $|q^{2}|$ one obtains
\begin{align}\label{DC67}
\lim_{q^{2}\rightarrow \infty}\lambda_{-}=
\begin{cases}
q^{2}&\text{for $Z>1$} \\
Zq^{2}&\text{for $Z<1$}\;,
\end{cases} \nn\\
\lim_{q^{2}\rightarrow -\infty}\lambda_{-}=
\begin{cases}
Zq^{2} &\text{for $Z>1$}\\
q^{2} &\text{for $Z<1$}\;.
\end{cases}
\end{align}
For $Z<Z_{c}$ the function $\lambda_{-}(q^{2}) $ remains negative and real for the range of real $q^{2}$ with $-\mu^{2}< q^{2}<0$ .

In contrast, for $Z>Z_{c}$ the graviton propagator $\lambda_{-}^{-1}(q^{2})$ has a second pole at $q^{2}=-\mu^{2}$, with a behavior for $q^{2}$ near $-\mu^{2}$ given by
\be\label{DC68}
\lambda_{-}(q^{2})=B_{t}(q^{2}+\mu^{2})\;,
\ee
where
\be\label{DC69}
B_{t}=\dfrac{1}{2}\Bigl(Z+1+\dfrac{A_{t}m^{2}}{m_{t}^{2}}\Bigr)=\dfrac{Zy}{y-Z(1-y)}\;.
\ee
For $Z>Z_{c}$ one has $m_{t}^{2}<0$ and $B_{t}<0$. The negative prefactor of $q^{2}+\mu^{2}$ corresponds to a negative residuum at the pole at $q^{2}=-\mu^{2}$. This indicates a ghost instability. A stable theory with a well behaved graviton propagator for  all momenta therefore requires the ``stability condition"~\eqref{DC63}, $Z<Z_{c}$.

With this condition the graviton propagator has only a single pole in the complex $q^{2}$- plane at $q^{2}=0$. In the standard normalization with dimensionless metric or vierbein fields the graviton propagator is given by
\be\label{DC69A}
G_\textup{grav}(q^{2})=\dfrac{4}{m^{2}\lambda_{-}(q^{2})}\;.
\ee
This is the graviton propagator~\eqref{IN1} mentioned in the introduction.
For $q^{2}$ near zero it reads, cf.~eq.~\eqref{DC61},
\be\label{DC70}
G_\textup{grav}(q^{2})=\dfrac{4}{M^{2}q^{2}}\Bigl(1+(Z-y)\dfrac{q^{2}}{M^{2}}\Bigr)^{-1}\;.
\ee
While the truncated propagator has a ghost pole at \mbox{$q^{2}=-M^{2}/(Z-y)$,} this pole is not present in the propagator~\eqref{DC69A}.
It is therefore an artefact of the truncation~\citep{PLCW}.

\subsection{Massive spin-two particle\label{subsec: Massive spin-two particle}}
\smallskip

In the stable range for $Z<Z_{c}$ an approximate form for $\lambda_{+}(q^{2})$ in a range of momenta in the vicinity of the pole at $q^{2}=-\mu^{2}$ is given by
\be\label{DC71}
\lambda_{+}=B_{t}(q^{2}+\mu^{2})\;.
\ee
With $m_{t}^{2}>0$ one infers $B_{t}>0$. The eigenmode $\lambda_{+}(q^{2})$ corresponds to a stable massive spin-two particle, with mass $\mu$ and positive residuum at the pole $B_{t}^{-1}>0$. This resembles to similar findings in refs~\cite{HS1,HS2, HS3, SVN, NARR, NIRRU}.

In the stable range both eigenvalues of the inverse propagator matrix~\eqref{DC56} correspond to stable modes. There is no instability in this sector. At $q^{2}=0$ one finds a positive value of $\lambda_{+}$, 
\be\label{DC72}
\lambda_{+}(q^{2}=0)=m^{2}-M^{2}\;.
\ee
For the boundary case $M^{2}=m^{2}$ one observes a second stable massless particle, with
\be\label{DC73}
\lambda_{+}=Zq^{2}\;,\quad \lambda_{-}=q^{2}\;.
\ee
In this limit $Z_{c}$ diverges. A future exploration of the limit with two massless tensor fields is conceptually rather interesting, but not the focus of this paper.

The condition of stability $Z<Z_{c}$ corresponds to a lower bound for $\mu^{2}$,
\be\label{143AA}
\mu^{2}>m^{2}(1-y)^{2}=\frac{(1-y)^{2}}{y}M^{2}   \;,
\ee
where we recall eq.~\eqref{DC67}. Except for the limiting case $y\rightarrow 1$ the heavy mass is at least of the order $M$, such that the massive spin-two particle induces only tiny corrections for low energy phenomenology.

If for a given $m^{2}$ and $Z$ one starts at $M^{2}=0$ and switches on $M^{2}$ continuously, the double pole at $q^{2}=0$ for $M^{2}=0$ first turns to a ghost instability since the condition~\eqref{DC63} will be violated for small enough $y$. Once $y$ reaches the critical value saturating eq.~\eqref{DC63} the pole jumps from $\lambda_{-}$ to $\lambda_{+}$ and the sign of the residuum changes correspondingly. This explains why the massive spin-two particle cannot be found as a small deformation in the vicinity of a massless graviton. The massive particle pole is outside the range of validity of a polynomial expansion in $q^{2}$ for the inverse propagator.

\subsection{Analyticity\label{subsec: Analyticity}}
\smallskip

Analytic continuation can be discussed on different levels. On an overall level one has to specify how all fields are analytically continued. This includes the gauge fields, such that the gauge group $SO(4)$ is analytically continued to the non-compact gauge group $SO(1,3)$. We will describe this procedure in a separate paper. In the present paper we discuss analytic continuation on the level of propagators in flat space. This amounts to a discussion of the behavior of propagators in the plane of complex $q^{2}$ or complex $q_{0}$. Analytic continuation interpolates from the euclidean relation $q^{2}=q^{\mu}q_{\mu}=q_{0}^{2}+\vec{q}^{\;2}$ to the Minkowski relation $q^{2}=-q_{0}^{2}+\vec{q}^{\;2}$. It requires that there exists a continuous path between the two limits in the complex plane that is not obstructed by non-analyticities as poles or branch cuts.

For an overall picture of the $q^{2}$-dependence of the two eigenvalues $\lambda_{\pm}(q^{2})$ we further note that at a critical value $q_{c}^{2}<0$ both eigenvalue can coincide at branch points $q_{c}^{2}$,
\be\label{DC74}
\lambda_{+}(q_{c}^{2})=\lambda_{-}(q_{c}^{2})\;.
\ee
This occurs when the square root in eq.~\eqref{DC60} vanishes, determining
\be\label{DC75}
\dfrac{q_{c\pm}^{2}}{m^{2}}=-\dfrac{1-y}{(1-Z)^{2}}\Bigl{\lbrace}Z+1-2y\mp2\sqrt{1-y}\;\sqrt{Z-y}\Bigr{\rbrace}\;.
\ee
The condition for the existence of real intersection points is $Z\geqslant y$. For $Z=y$, which belongs to the stable region, $\lambda_{+}(q^{2})$ and  $\lambda_{-}(q^{2})$ touch each other at $q_{c}^{2}=-m^{2}$. For $Z>y$ one has a finite region $q_{c_{-}}^{2}<q^{2}<q_{c_{+}}^{2}$ for which the argument in the square root of eq.~\eqref{DC60} becomes negative. In this region $\lambda_{+}(q^{2})$ and $\lambda_{-}(q^{2})$ have a non-vanishing imaginary part for real $q^{2}$. (For $Z\rightarrow 1$ one has $q_{c_{-}}^{2}\rightarrow -\infty\;, \; q_{c_{+}}^{2}\rightarrow -m^{2}/4$, while for $Z\neq 1$, $Z>y$ both $q_{c_{-}}^{2}$ and $q_{c_{+}}^{2}$ are finite.) We observe that $Z>y$ is compatible with $Z<Z_{c}=y/(1-y)$ for a suitable range of $Z$, and we focus first on this case. 

The region with an imaginary part of $\lambda_{\pm}(q^{2})$ occurs always beyond the location of the second pole of the propagator, $q_{c_{\pm}}<-\mu^{2}$. This is visible from
\be\label{DC76}
\dfrac{q_{c}^{2}}{m^{2}}=-\dfrac{\mu^{2}}{m^{2}}-x_{t}\;,
\ee
with
\be\label{DC77}
\!\!\!\!\!\!\!\!x_{\pm}\!\!=\dfrac{1}{(Z-1)^{2}}\Bigl(b\mp \sqrt{b^{2}-(Z-1)^{2}v^{2}}\Bigr)\,,\quad\!\! v=\dfrac{m_{t}^{2}}{m^{2}}\,,\!\!\!\!\!\ee
where
\be
\begin{split}
 b&=2(1-y)^{2}+(Z-1)\Bigl(1-y-\dfrac{\mu^{2}}{m^{2}}\Bigr)\\
&=(1-y)\Bigl[(Z+1)(1-y)+\dfrac{y}{Z}(Z-1)^{2}\Bigr]\;.
\end{split}
\ee
The argument of the square root being smaller than $b^{2}$, both $x_{+}$ and $x_{-}$ are positive since $b>0$.

In the complex $q^{2}$- plane the graviton propagator has a branch cut for real negative $q^{2}$, extending from $q_{c_{-}}^{2}$ to $q_{c_{+}}^{2}$. Except for the pole at $q^{2}=0$ and this cut, the graviton propagator is analytic, decaying~$\sim |q|^{-2}$ for large $|q|$. Nothing obstructs analytic continuation from euclidean space to Minkowski space. We can therefore extend our analysis to a Minkowski metric, $g_{\mu\nu}=\eta_{\mu\nu}$ and $q^{2}=-q_{0}^{2}+\vec{q}^{\,2}$. In the complex $q_{0}$ - plane the graviton propagator has two poles at $q_{0}=\pm\sqrt{\vec{q}^{\,2}}$. It inherits the cuts on the real $q_{0}$ - axis, extending from  $q_{0}^{2}=|q_{c_{-}}^{2}|+\vec{q}^{\,2}$ to $|q_{c_{+}}^{2}|+\vec{q}^{\,2}$. The prescription for the usual infinitesimal $i\varepsilon$ -  terms is dictated by analytic continuation. This graviton propagator is well behaved in Minkowski space, without any instability.

For the particular case $Z=y$ one has the particularly simple behavior
\be\label{AC1}
\lambda_{-}=\frac{M^{2}}{m^{2}}q^{2}\;,\quad 
\lambda_{+}=q^{2}+m^{2}-M^{2}\;.
\ee
For $q_{c}^{2}=-m^{2}$ the two eigenvalues are equal, $\lambda_{-}=\lambda_{+}=-M^{2}$.
Finally, for $Z<y$ the branch points move to complex $q^{2}$, such that some care will be needed for analytic continuation.

In summary of this section both stability conditions~\eqref{DC59}~and~\eqref{DC63} are necessary for a stable graviton sector. If realized, the massless graviton is accompanied by a massive spin two particle which is stable as well. This is an example how a pair of massless and massive spin two particles can be obtained in a rather simple setting. No problem of consistency or stability is visible in this sector.

\section{Stability of flat space\label{sec: Stability of flat space}}

In the presence of $L_{R}$ flat space cannot be the minumum of the euclidean effective action. For flat space one has \mbox{$\Gamma=0$.} The curvature scalar $R$ can take positive and negative values, however. Since the term $L_{R}$ is dominant for small momenta, positive $R$ and therefore negative $L_{R}$ lead to $\Gamma<0$. This issue is well known in Einstein gravity. In Einstein gravity it leads to an unbounded Einstein-Hilbert action such that the euclidean functional integral is not well defined. For our model of pregeometry the functional integral based on the classical action is well defined. This extends to the high momentum behavior of the effective action, and therefore the behavior for large $|R|$, which are well behaved. The presence of $L_{R}$ in the effective action indicates, however, that flat space is a saddle point of the euclidean effective action, but not a minimum. 
The fact that flat space is not a minimum of the euclidean effective action does not imply instability of Minkowski space. A good example is again the Einstein-Hilbert action. Despite possible negative values of the effective action Minkowski space is stable due to the positive energy theorem. What is required for stability is the absence of tachyonic modes. This issue will be discussed next.

\subsection{Scalar propagators\label{subsec: Scalar propagators}}
\smallskip

In the scalar sector one has three physical scalar modes, namely $\sigma$, $\tilde{w}$, $\tilde{v}$. The scalar $u$ is a gauge mode. The scalar $\tilde{v}$ does not mix with the other two physical scalars. 
It describes a massive particle with a stable propagator
\be\label{SP01}
G_{\tilde{v}}\sim (Zq^{2}+m^{2}+2M^{2})^{-1}\;,
\ee
as obtained from the action
\be\label{SP02}
S_{\tilde{v}}=\int_{q}v(-q)\dfrac{3q^{2}}{8}(Zq^{2}+m^{2}+2M^{2})v(q)\;.
\ee
The squared mass $m^{2}+2M^{2}$ is increased by $L_{R}$.

In the $\sigma-\tilde{w}$ sector one finds the inverse propagator matrix
\be\label{SP03}
\!\!\!\!\!\!P^{(\sigma\tilde{w})}\!(q^{2})\!=\!\dfrac{1}{3}\!\!\begin{pmatrix}
\!\!\tilde{Z}q^{4}\!+\!(\tilde{m}^{2}\!\!+\!2M^{2})q^{2}&-\Bigl(\dfrac{1}{2}\tilde{m}^{2}\!\!+\!M^{2}\Bigr)q^{2}\phantom{\Bigg{|}}\!\!\\-\Bigl(\dfrac{1}{2}\tilde{m}^{2}\!\!+\!M^{2}\!\Bigr)q^{2}&\dfrac{\tilde{m}^{2}}{4}q^{2}
\end{pmatrix}\!\!
.\!\!\!\!\!\!
\ee
The addition of the contribution $L_{G}$ changes $Z\to \tilde{Z}$, $m\2\to\tilde{m}\2$ according to eq.~\eqref{112A}.

The zero eigenvalues of $P^{(\sigma\tilde{w})}$ occur for 
\be\label{SP05}
q^{2}=0\;,\quad q^{2}=\nu^{2}=\dfrac{2M^{2}}{\tilde{Z}}(1+2\tilde{y})\;, \quad \tilde{y}=\dfrac{M\2}{\tilde{m}\2}
\;.
\ee
If both $\tilde{Z}$ and $\tilde{y}$ are positive, the second pole of the scalar propagator is located at positive $q^{2}$, in contrast to \mbox{$q^{2}=-\mu^{2}$} for the graviton sector. In the absence of $L_{G}$ this leads to a tachyonic instability, since in this case $\tilde{Z}=Z$ , $\tilde{y}=y$ , and both $Z$ and $y$ have to be positive for a stable graviton. We conclude that stability of flat space requires either $\tilde{Z}$ or $\tilde{m}\2$ or both to be negative, such that either $C$ or $n\2$ need to be negative. Stability of flat space requires
\be 
\label{158A}
\tilde{Z}<0\;,\quad \tilde{y}>-\dfrac{1}{2}\;,
\ee
or
\be 
\label{158B}
\tilde{Z}>0\;,\quad \tilde{y}<-\dfrac{1}{2}\;.
\ee
We will choose the first condition~\eqref{158A}.

The discussion of stability in the scalar sector is somewhat complex since different regions in the space of couplings $\tilde{Z}$, $\tilde{m}^{2}$ have to be distinguished. We first start with $\tilde{m}^{2}<0$ and turn later to positive $\tilde{m}^{2}>0$.
Restoring the normalization to scalar fields with dimension mass the corresponding renormalized propagator matrix becomes for $\tilde{m}^{2}<0$
\be\label{SP04}
\!\!\!\!\!\!P_{R}^{(\sigma\tilde{w})}(q^{2})=\dfrac{1}{3}\!\!\begin{pmatrix}
\tilde{Z}q^{2}\!\!+\!\tilde{m}^{2}\!\!+\!2M^{2}&\!\!\!,&-(\tilde{m}^{2}\!\!+\!2M^{2})\dfrac{q}{|\tilde{m}|}\\\!\!-(\tilde{m}^{2}\!\!+\!2M^{2})\dfrac{q}{|\tilde{m}|}&\!\!\!,&\!\!-q^{2}
\end{pmatrix}
\!\!
.\!\!\!\!\!\!
\ee
Metastability is expected to become visible in the behavior of one of the eigenfunctions for $q^{2}\rightarrow 0$. According to the effective low energy theory being general relativity, the physical scalar fluctuation in the metric should have the ``wrong" sign of the inverse propagator.

For general $q^{2}$ the two eigenvalues of $P_{R}$ are given by
\ba\label{SP06}
&\lambda_{\pm}(q)^{2}=\dfrac{1}{6}\Biggr{\lbrace}(\tilde{Z}-1)q^{2}+\tilde{m}^{2}(1+2\tilde{y})\\
&\quad\quad\pm\sqrt{\Bigl[(\tilde{Z}+1)q^{2}+\tilde{m}^{2}(1+2\tilde{y})\Bigr]^{2}\!\!-4(1+2\tilde{y})^{2}\tilde{m}^{2}q^{2}}\,\Biggr{\rbrace}\;. \nn
\end{align}
For $q^{2}$ near zero one obtains for $\tilde{m}^{2}<0$
\be\label{160}
\lambda_{+}(q^{2})=\dfrac{2\tilde{y}}{3}q^{2}+\dots\quad .
\ee
For $\tilde{m}\2 <0$ , $\tilde{y}<0$ the coefficient of $q\2$ is negative, in contrast to the opposite sign of $\lambda_{-}=yq^{2}$ for the graviton sector. This is the expected behavior for the Einstein-Hilbert action.
Around $q\2=0$ the second eigenvalue is negative
\be 
\label{160A}
\lambda_{-}(q\2)=\dfrac{1}{3}\Big{[}\tilde{m}\2(1+2\tilde{y})+(\tilde{Z}-1-2\tilde{y})q\2 +\dots\Big{]}\;.
\ee

At the other pole of the propagator one finds for $1+2\tilde{y}>2\tilde{y}/\tilde{Z}$ that it corresponds again to $\lambda_{+}$
\ba 
\label{160B}
&\lambda_{+}(q\2=\nu\2)=0\;,\nn\\
 &\lambda_{-}(q\2=\nu\2)=\dfrac{\tilde{m}\2}{3}(1+2\tilde{y})\Big{(}1+2\tilde{y}-\dfrac{2\tilde{y}}{\tilde{Z}}\Big{)}\;.
\end{align}
On the other hand, for $1+2\tilde{y}<2\tilde{y}/\tilde{Z}$  the role of $\lambda_{+} $ and $\lambda_{-}$ is exchanged. For $\tilde{m}\2 <0$ , $1+2\tilde{y}>2\tilde{y}/\tilde{Z}$ both poles correspond to zeros of $\lambda_{+}$, while $\lambda_{-}$ remains negative for all $q\2$, without any zero. For $\tilde{m}\2 <0$ , $1+2\tilde{y}<2\tilde{y}/\tilde{Z}$  the pole at $q\2=0$ belongs to $\lambda_{+}$, while the pole at $q\2=\nu\2$ occurs for $\lambda_{-} $ .
Expanding the eigenvalues in $q\2-\nu\2$ one obtains for $1+2\tilde{y}>2\tilde{y}/\tilde{Z}$
\ba 
\label{160C}
&\dfrac{\partial\lambda_{+}}{\partial q\2}\,\rule[-3mm]{0.1mm}{5mm}_{\: q\2=\nu\2}=-\dfrac{2\tilde{y}}{3}\big{(}1+2\tilde{y}-\dfrac{2\tilde{y}}{\tilde{Z}}\Big{)}^{-1}>0\;,\\
&\dfrac{\partial\lambda_{-}}{\partial q\2}\,\rule[-3mm]{0.1mm}{5mm}_{\: q\2=\nu\2}=\dfrac{1}{3}(\tilde{Z}-1)+\dfrac{2\tilde{y}}{3}
\big{(}1+2\tilde{y}-\dfrac{2\tilde{y}}{\tilde{Z}}\Big{)}^{-1}<0\;.\nn
\end{align}
The residuum at the pole at $q\2=\nu\2$ is therefore positive. This corresponds to a stable particle with mass $\nu$. For $1+2\tilde{y}<2\tilde{y}/\tilde{Z}$ the role of $\lambda_{+} $ and $\lambda_{-}$ are interchanged. The residuum at the pole at $q\2=\nu\2$  is now negative, indicating a ``ghost-excitation" with a negative kinetic term.

We can perform a similar discussion for positive $\tilde{m}\2>0$. The element in the lower right corner of the matrix~\eqref{SP04} is now $+q\2$ instead of $-q\2$ .
Correspondingly, the two eigenvalues read
\ba 
\label{160D}
&\lambda_{\pm}(q\2)=\dfrac{1}{6}\bigg{\lbrace} (\tilde{Z}+1) q\2+\tilde{m}\2(1+2\tilde{y})\\
&\quad\pm \sqrt{\big{[}(\tilde{Z}-1) q\2+\tilde{m}\2(1+2\tilde{y})\big{]}^{2}+4(1+2\tilde{y})\2\: \tilde{m}\2 q\2} \; \bigg{\rbrace}\;.\nn
\end{align}
The pole at $q\2=0$ corresponds now to $\lambda_{-}$, with expansion
\be 
\label{160E}
\lambda_{-}(q\2)=-\dfrac{2\tilde{y}}{3}q\2+\dots\;.
\ee
With $\tilde{y}>0$ the coefficient of $q\2$ is again negativ. For $1+2\tilde{y}+2\tilde{y}/\tilde{Z}>0$ the second pole at $q\2=\nu\2$ occurs for $\lambda_{-}$ as well, while
\be 
 \label{160F}
\lambda_{+}(q\2=\nu\2)=\dfrac{1}{2}(1+2\tilde{y})\Big{(}1+2\tilde{y}+\dfrac{2\tilde{y}}{\tilde{Z}}\Big{)}>0\;.
\ee
In this case $\lambda_{+}$ is not related to a propagating degree of freedom. The function $\lambda_{+}^{-1}(q\2)$ does not have any pole in the complex $q\2$-plane. With
\ba 
\label{160G}
&\dfrac{\partial\lambda_{-}}{\partial q\2}\big{(} q\2=\nu\2\big{)}=\dfrac{2\tilde{y}}{3}\big{(}1+2\tilde{y}+\dfrac{2\tilde{y}}{\tilde{Z}}\Big{)}^{-1}\;,\nn\\
&\dfrac{\partial\lambda_{+}}{\partial q\2}\big{(}q\2=\nu\2\big{)}=\dfrac{\tilde{Z}+1}{3}-\dfrac{2\tilde{y}}{3}
\big{(}1+2\tilde{y}+\dfrac{2\tilde{y}}{\tilde{Z}}\Big{)}^{-1}\;,
\end{align}
we find at $q\2=\nu\2$ a positive value for $\partial\lambda_{-}/\partial q\2$ , indicating a stable particle. For $
 1+2\tilde{y}+2\tilde{y}/\tilde{Z}<0$ the role of $\lambda_{+} $ and $\lambda_ {-}$ are interchanged. The pole at $q\2=\nu\2$ occurs now for $\lambda_{+}$ , with $\partial\lambda_{+}/\partial q\2<0$ . The pole at $q\2=\nu\2$ corresponds to a ghost. 
 
 In summary, for $ \tilde{Z}<0$ , $\tilde{y}>-1/2$ there is no tachyonic instability. A stable massive particle can occur for both positive or negative values of $\tilde{m}\2$ and $\tilde{y}$ , provided
 \be 
 \label{160H}
1+2\tilde{y}+\dfrac{2|\tilde{y}|}{ \tilde{Z}}>0
\;.
\ee
If the condition~\eqref{160H} is violated, the massive pole corresponds to a ghost. For the pole at $q\2=0$ the coefficient is always negative, indicating for euclidean signature a saddle point, in complete analogy to the Einstein-Hilbert action for general relativity.
This reflects the observation that the configuration with $e_{\mu}^{m}=\delta_{\mu}^{m}$, $A_{\mu mn}=0$ cannot be the minimum of the euclidean action. The term $L_{R}$ linear in $F$ is not positive definite and can take negative values. The flat space configuration with vanishing gauge fields is therefore a saddle point. 

\subsection{Metastability of flat euclidean space\label{subsec: Metastability of flat euclidean space}}
\smallskip

The euclidean action based on $L_{U}+L_{F}$ involves two squares of real tensors. Since we restrict the vierbein to values for which the determinant $e$ is positive, $e\geqslant 0$, an effective action containing only the terms $L_{U}+L_{F}$ is positive definite, $\Gamma\geqslant 0$. The minimum occurs for $\Gamma=0$. This is realized by flat space and zero gauge fields.
Adding the invariant $L_{R}$, the flat space solution with vanishing gauge fields becomes a saddle point. As argued before, this follows from the simple fact that $F=F_{\mu\nu}^{\quad\mu\nu}$ can take positive and negative values. For every $M^{2}\neq 0$ there are therefore field configurations for which the action becomes negative. Flat space remains a solution, but is no longer a minimum of the euclidean action. It is turned to a saddle point. For $M^{2}>0$ the graviton sector is  stable, i.e. the action increases to positive values $\Gamma>0$ for nonzero inhomogeneous small $t_{\mu\nu}$ or $E_{\mu\nu}$. In contrast, the scalar direction has partially the opposite property. For small $|q\2 |$ the action can decrease for nonzero small values of the scalar fields $\sigma$ and $\tilde{v}$. 

For larger momentum also the term $\sim L_{G}$ plays an important role. For the graviton sector it only replaces $Z$ by $Z+B$ . Thus the graviton sector is unaffected for our choice $B=0$ . In contrast, the scalar sector depends strongly on $\tilde{Z}$. Omitting $L_{G}\,$ (taking $\tilde{Z}=Z , \, \tilde{y}=y )$ one would find a tachyonic instability in the scalar sector. This is avoided for $\tilde{Z}<0$ . On the other hand, $\tilde{Z}<0$ implies in the derivative expansion considered here an unbounded euclidean action. This is apparent if we decompose the squared field strength as
\be\label{SP10}
\!\!\!\!\dfrac{Z}{8}F_{\mu\nu\rho\sigma}F^{\mu\nu\rho\sigma}=\alpha F^{2}\!\!+\!\beta\tilde{F}_{\mu\nu}\tilde{F}^{\mu\nu}\!\!+\!\gamma W_{\mu\nu\rho\sigma} W^{\mu\nu\rho\sigma}\,,\!\!\!\!
\ee
with
\be\label{SP11}
\quad \quad\tilde{F}_{\mu\nu}=F_{\mu\nu}-\dfrac{1}{4}Fg_{\mu\nu}\,,\quad\quad\quad \tilde{F}^{\mu\nu}g^{\mu\nu}=0\,, 
\ee
and
\begin{align}\label{179A}
&W_{\mu\nu\rho\sigma}=F_{\mu\nu\rho\sigma}-\dfrac{1}{2}(g_{\mu\rho}F_{\nu\sigma}+g_ {\nu\sigma}F_{\mu\rho}-g_{\mu\sigma}F_{\nu\rho}-g_{\nu\rho}F_{\mu\sigma}) \nn\\
&\quad\quad\quad\quad\quad\quad\quad +\dfrac{1}{6}F(g_{\mu\rho}g_{\nu\sigma}-g_{\mu\sigma}g_{\nu\rho})\; , 
\end{align}
where
\be
 W_{\mu\nu\rho\sigma}g^{\nu\sigma}=W_{\mu\nu\rho\sigma}g^{\mu\rho}=0\;.
\ee
The coefficients are given by
\be\label{SP12}
\alpha=\dfrac{Z}{48} ,\quad\beta=\dfrac{Z}{4} , \quad\gamma=\dfrac{Z}{8}\;.
\ee
One finds for the sum
\ba 
\label{167}
L_{F}+&L_{R}+L_{G}=\dfrac{Z}{8}W_{\mu\nu\rho\sigma}W^{\mu\nu\rho\sigma}+\dfrac{Z+2B}{4	}\tilde{F}_{\mu\nu}
\tilde{F}^{\mu\nu}\nn\\
&+\dfrac{1}{48}\big{(}2\tilde{Z}-(Z+2B)\big{)}F\2-\dfrac{M\2}{2}F\;.
\end{align}
Here we generalize for $B\neq 0$
\be 
\label{168}
\tilde{Z}=Z+4B+12C\;,
\ee
such that our discussion of the scalar sector remains valid for $B\neq 0$. For $\tilde{Z}<0$ and $Z+2B>0$ the coefficient of the term $\sim F\2$ is negative, implying an unbounded euclidean action.

\subsection{Stability of Minkowski space\label{subsec: Stability of Minkowski space}}
\smallskip

As we have discussed above, no tachyonic or ghost instability occurs in the graviton and scalar sectors for the parameter range
\ba 
\label{168A}
&m\2>0\;,\quad y>0\;,\quad 0<Z<\dfrac{y}{1-y}\;,\nn\\
&\tilde{y}>-\dfrac{1}{2}\;,\quad \tilde{Z}<-\dfrac{2|\tilde{y}|}{1+2\tilde{y}}\;.
\end{align}
We will see below that these conditions ensure also the stability of the other fluctuation modes. The status of ghost instabilities is not completely understood. If one only wants to avoid tachyonic instabilities the conditions on $Z$ and $\tilde{Z}$ get relaxed to $Z>0$, $\tilde{Z}<0$.

With parameters in the range~\eqref{168A} this suggest stability of Minkowski space. This property of the field equations becomes more manifest if we add in our setting a further scalar field $\phi$ with effective action
\be 
\label{168B}
\Gamma_{\phi}=\int_{x}e\:\delta(\phi\2-F)\2\;,
\ee
with
\be 
\label{168C}
\delta=-\dfrac{1}{48}\big{(}2\tilde{Z}-(Z+B)\big{)}>0\;.
\ee
The field equations for the extended system are equivalent to the original system if we discard solutions with $\phi=0$ . Indeed, inserting the solution $\phi\2=F$ for the field equation for $\phi$ into the effective action the contribution $\Gamma_{\phi}$ vanishes. With our choice of $\delta$ the coefficient $\sim F\2$ of the combination $L_{F}+L_{G}+L_{\phi}$ vanishes. We end in the scalar sector with the expression
\be 
\label{168D}
L_{\phi}'=\int_{x} e\bigg{\lbrace}-\Big{(}\dfrac{M\2}{2}+2\delta\phi\2\Big{)}F+\delta\phi^{4}\bigg{\rbrace}\;.
\ee
Replacing $F$ by $R$ the resulting theory of general relativity coupled to a scalar field is well known to be stable.

A Weyl scaling to the Einstein frame induces a kinetic term for $\phi$. The scalar field $\phi$ is stable, with effective potential in the Einstein frame
\be 
\label{168E}
V_{E}=\delta\overline{M}^{4}\phi^{4}(M\2+4\delta\phi\2)^{-2}\;.
\ee
Here $\overline{M}$ is the observable Planck mass. The situation is analogous to general relativity with higher derivative invariants. Stability of Minkowski space requires the coefficient of the terms $\sim R\2$ in the effective action to be negative, similar to the term $\sim -\delta F\2$ for pregeometry. 
In the version with an additional scalar field $\phi$ the scalar sector of our model of pregeometry ressembles closely general relativity with an additional stable scalar field.

The action for the euclidean metric of Einstein gravity exhibits also a saddle point, with scalar fluctuations around flat space leading to a lowering of the euclidean action. At least for the analytically continued model in Minkowski space this is actually not a sign of instability. Due to the positive energy condition Minkowski space is stable in Einstein gravity. The scalar mode is actually not a propagating mode. Combined with the necessary projectors the scalar propagator turns out to be given by $(\vec{q}^{\;2})^{-1}$ instead of $(q^{2})^{-1}$~\cite{CWMF}. There is therefore no pole for $\vec{q}^{\;2}\neq 0$. The scalar propagator accounts for Newton's potential, rather than being a physical propagating particle. We expect a similar behavior for our model, at least for the region of momenta $q^{2}\ll M^{2}$. The pole at $q^{2}=0$ in the stability analysis above is expected to correspond to the physical scalar in the metric for Einstein gravity. The other pole at $q^{2}=\nu^{2}$ can be associated to the stable scalar field $\phi$.

\subsection{Classical action and effective action\label{subsec: Classical action and effective action}}
\smallskip

The metastability of euclidean flat space and the stability of Minkowski space points to an important difference between the classical action and the effective action. For the classical action all field configurations are included in the functional integral. A lack of boundedness from below obstructs a well defined functional integral. For the effective action we are concerned with the stability of solutions of the field equations derived from it. Not all arbitrary field configurations are reached by such solutions. Einstein's equation imply constraints that imply that the scalar part in the metric cannot be chosen arbitrarily. It is a generalization of Newton's potential which is determined as a fixed functional of the other fluctuations. For Minkowski signature this is the basic reason why the apparent metastability of flat space does not induce any instability of the solutions of field equations. One concludes that on the level of the effective action not every instability in the space of all arbitrary field configuration leads to an instability of the field configurations that can be reached by solutions of field equations.

It seems rather likely that these features also play a role for the effective action in a euclidean setting. The field configurations that can be reached by solutions of the field equations include the presence of conserved sources and arbitrary boundary conditions. Nevertheless, the metastability for arbitrary configurations in the scalar sector may not lead to metastable solutions of field equations. We conclude that unboundedness of the euclidean effective action points to a possible problem, but is not yet per se and indication for an unacceptable instability.

If one formulates the classical or microscopic action for the definition of the functional integral at distances much smaller than $m^{-1}$, $|\tilde{m}|^{-1}$, there is not a very direct relation to the quantum effective action for momenta $q^{2}$ of the order of $m^{2}$, $\tilde{m}^{2}$ or $M^{2}$ .
By virtue of universality a large family of classical actions can lead to the same effective action in this momentum range. Even if we associate the effective action at some very short distance scale $l_{0}$ with the classical action, the effective couplings will change their values between $q^{2}\approx l_{0}^{-2}$ and  $q^{2}\approx M^{2}$. This flow of couplings can be a very substantial effect if $M^{2} l_{0}^{2}\ll 1$ . 

The stability discussion in the sector of scalar fluctuations involves the couplings at $q^{2}\approx M^{2}$. A tachyonic instability can be avoided if the running coupling $\tilde{Z}\big{(}q^{2}\approx M^{2}\big{)}$ is negative. This is enough for a stable particle pole at $q\2 =\nu\2\approx O(M\2)$, $\nu\2 <0$. So far we have not discussed a possible running of the couplings. Constant couplings may be a valid approximation for the momentum range which is relevant for the stability discussion. In general, however, the dimensionless couplings will be functions of $q\2 / k\2 $, where $k$ is some suitable renormalization scale. In a covariant setting $q\2$ stands for squared covariant derivatives, such that the flow of couplings is associated with an effective field dependence. It is well conceivable that couplings $\tilde{Z}$ and $\tilde{m}\2 $ flow for $q\2 \to \infty$ to positive values. This may well lead to a bounded euclidean effective action, in correspondence to the bounded classical action. The boundedness would become visible for $q\2 \gg m\2 , |\tilde{m}|\2$, $M\2$ or very large field values. Nevertheless, flat space would not be the minimum of the euclidean effective action but rather be a saddle point.

\subsection{Stability of other modes\label{subsec: Stability of other modes}}
\smallskip

 So far we have found a region~\eqref{168A} in parameter space for which the traceless transversal tensor and scalar fluctuation modes are stable. We will conclude this section by establishing that for this parameter range also all other fluctuation modes are stable. For simplicity we concentrate on $B=0$ . We also omit the gauge modes, setting $l_{\mu}=0$, $\kappa_{\mu}=0$, $M_{\mu\nu}=0$, $u=0$ . 
 We begin with the gauge boson fluctuation $C_{\mu\nu \rho}$ . In quadratic order in an expansion around flat space there are no contributions from $L_{G}$ and the inverse propagator is proportional
 \be
 \label{168F}
P_{C}=\frac{Z}{2}\big{(}q\2 + \mu_{C}\2  \big{)}\;,\quad \mu_{C}\2 = \frac{(1+2y) m\2}{Z}\; .
\ee
This fluctuation has the same mass as the scalar fluctuation $\tilde{v}$ and is therefore stable.

The transverse vectors $\gamma_{\mu}$ and $w_ {\mu}^{(t)}$ are mixed. The inverse propagator matrix, 
\be
\label{168G}
\!\!\!\!P_{w\gamma}=\frac{2}{9}
\begin{pmatrix}
Zq^{2}+4m^{2}+2n\2 +M\2\!\! \!\!&&\dfrac{3}{2}  q\2 n\2 M\2 \\
\dfrac{3}{2} q\2 n\2 M\2\!\!\!\! &&\phantom{\Bigg|} \; \dfrac{9}{8}q^{4}(m\2 + n\2 ) 
\end{pmatrix}\!\!,\!\!\!\!
\ee
implies for this sector a massless fluctuation with pole at $q\2=0$, and a massive excitation with pole at $q\2=-\mu_{w\gamma}\2 $ ,
\be
\label{168H}
\mu_{w\gamma}\2 =\frac{4m^{4}+6m\2 n\2 + M\2 (m\2 -3n\2 -2M\2 )}{Z(m\2 + n\2 )}
\;.
\ee
For a whole range of $n\2/m\2$ and $M\2 / m\2$ the massive particle is stable, $\mu_{w\gamma}\2 >0$ .

Finally, for the sector of the fluctuations $v_{\;\mu} ^{(t)} $ and $b_{\mu\nu}$ we employ
\be
\label{168I}
\tilde{b}^{ \mu}=\varepsilon^{\mu\nu\rho\sigma} \partial_ {\nu} b_{\rho\sigma}\;,\quad \partial_{\mu}\tilde{b}^{\mu}=0\;.
\ee
The transverse vectors $v_{\;\mu}^{(t)}$ and $\tilde{b}_{\mu}$ mix, with propagator matrix
\be
\label{168J}
P_{vb}=\frac{1}{2}\begin{pmatrix}
Zq\2 + m\2 +M\2 \phantom{\bigg|} && \dfrac{M\2 }{2}\\
\dfrac{M\2}{2}&& \dfrac{m\2}{2}
\end{pmatrix}
\;.
\ee
In this sector one finds again a stable massive particle with poisitive mass term 
\be
\label{168K}
\mu_{bv}\2 =\frac{\big{(}1+y-\frac{1}{2} y\2 \big{)} m\2}{Z}\;.
\ee

We conclude that no tachyonic fluctuations around flat space are present. For $U=0$ Minkowski space is a solution of the analytically continued model with Minkowski signature. The findings in the present section implies that the linearized field equations in the vicinity of Minkowski space have no unstable solution. In this respect our model of pregeometry shares the stability property of general relativity.

\section{Cosmological constant\label{sec: Cosmological constant}}

The issue of metastability of the flat space solution concerns partly the physics at low momenta and geometries with large volumes. If we add a cosmological constant $U$ in eq.~\eqref{EA1}, it plays a central role for the behavior for $q^{2}\rightarrow 0$. The cosmological constant $U$ does not vanish for $q^{2}\rightarrow 0$, in contrast to all other terms that involve derivatives. In this section we include in the effective action a small positive cosmological constant $U>0$. We will see its profound implications for the effective action for euclidean gravity. 

\subsection{Spheres with vanishing covariant derivative of vierbein\label{subsec: Spheres with vanishing covariant derivative of vierbein}}
\smallskip

Let us consider a particular family of field configurations for which the vierbein is covariantly conserved, while the metric describes a sphere with radius $L$, 
\begin{align}\label{CC1}
U_{\mu\nu}{}^{m}=0\;,\quad\quad F_{\mu\nu\rho\sigma}=R_{\mu\nu\rho\sigma}\;,\nn \\
R_{\mu\nu\rho\sigma}=\dfrac{1}{L^{2}}(g_{\mu\rho}g_{\nu\sigma}-g_{\nu\rho}g_{\mu\sigma})\;,  \nn \\
F_{\mu\nu}=R_{\mu\nu}=\dfrac{3}{L^{2}}g_{\mu\nu}\;, \quad F=R=\dfrac{12}{L^{2}}\;.
\end{align}
The effective action~\eqref{EA1} for this configuration is given for $B=0$ by
\be\label{CC2}
\begin{split}
\Gamma(L)=&cL^{4}\biggl(U-\dfrac{6M^{2}}{L^{2}}+\dfrac{3Z+72C}{L^{4}}\biggr) \\
=&c(UL^{4}-6M^{2}L^{2}+6\tilde{Z}-3Z)\;.
\end{split}
\ee
Here we have employed
\be\label{CC3}
\int_{x} e=cL^{4}\;.
\ee

The minimum of $\Gamma(L)$ obeys
\be\label{CC4}
UL_{0}^{2}-3M^{2}=0\;,
\ee
or
\be\label{CC5}
L_{0}^{2}=\dfrac{3M^{2}}{U}\;.
\ee
This corresponds to the solution of the Einstein equation
\be\label{CC6}
M^{2}(R_{\mu\nu}-\dfrac{1}{2}Rg_{\mu\nu})=-Ug_{\mu\nu}\;,\quad R=\dfrac{4U}{M^{2}}\;.
\ee
At the minimum, the effective action is negative,
\be\label{CC7}
\Gamma_{0}=-c(UL_{0}^{4}-3Z)=-c\biggl(\dfrac{9M^{4}}{U}-6\tilde{Z}+3Z\biggr)\;.
\ee
For a small ratio $U/M^{4}\ll 1$ the effective action $\Gamma_{0}$ takes large negative values, and the contributions $\sim \tilde{Z},\, Z$ can be neglected.

Nevertheless, the function $\Gamma(L)$ remains bounded from below even for very small nonzero $U/M^{4}$.
For $U>0$ flat space is no longer a solution of the field equations. Instead, we find a new solution of the field equations corresponding to the sphere~\eqref{CC1} with $L^{2}$ given by eq.~\eqref{CC5}.

\subsection{Minimum of euclidean effective action\label{subsec: Minimum of euclidean effective action}}
\smallskip

It is an interesting question if this solution corresponds to the minumum of the euclidean action~\eqref{EA1} on the space of possible solutions of the field equations. For this question we have to analyze fluctuations around this solution. We present several arguments suggesting that the solution~\eqref{CC1},~\eqref{CC5} could indeed be the minumum of the euclidean effective action on the space of possible solutions of field equations. 

We start by discussing the subspace of fluctuations for which the vierbein is covariantly conserved, $U_{\mu\nu}{}^{m}=0$. On this subspace the effective action becomes 
\ba\label{CC8}
\Gamma =& \int_{x} e \bigg{\lbrace} U-\dfrac{M^{2}}{2} R+\tilde{\alpha} R^{2}+\beta\tilde{R}_{\mu\nu}\tilde{R}^{\mu\nu}+\gamma C_{\mu\nu\rho\sigma}C^{\mu\nu\rho\sigma}\bigg{\rbrace}\;,  \nn\\
\tilde{\alpha}=& \frac{Z}{48}+\frac{C}{2}=\frac{1}{48}(2\tilde{Z}-Z)\;.
\end{align}
with $\tilde{R}_{\mu\nu}=R_{\mu\nu}-\dfrac{1}{2}Rg_{\mu\nu}$ and $C_{\mu\nu\rho\sigma}$ the Weyl tensor. The metric is the only relevant degree of freedom, such that the situation is precisely the same as for general relativity with effective action~\eqref{CC8}. We assume a form of $L_{G}$ for which the coefficients $\beta$ and $\gamma$ remain positive. With $e=\sqrt{g}$ eq.~\eqref{CC8} is a particular form of metric gravity. We may discuss the question of the minumum of $\Gamma$ for arbitrary couplings $U>0$, $M^{2}>0$, $\beta, \gamma>0$.

Consider first the limit $\beta\rightarrow \infty , \gamma\rightarrow \infty$. The minimum of $\Gamma$ has to obey in this limit $\tilde{R}_{\mu\nu}=0$, $C_{\mu\nu\rho\sigma}=0$. Solutions of the field equations with these properties are maximally symmetric spaces. On this subspace the spheres have lower action as compared to the ones with $R<0$. We have already found the minumum for the space of spheres. We conclude that for $\beta\rightarrow \infty , \gamma\rightarrow \infty$ the minimum of the effective action  on the space of solutions of the field equations is indeed given by a sphere with radius determined by eq.~\eqref{CC5}.

We can consider an alternative limit with finite $\beta$ and $\gamma$, taking now $U$, $M^{2}$ and $\tilde{\alpha}$ to zero. The effective action is positive semidefinite and its minimum occurs for maximally symmetric spaces. For spheres there is a flat direction corresponding to $L$. Allowing for $\tilde{\alpha}\neq 0$, the flat direction remains present, c.f. eq.~\eqref{CC2}. Adding $M^{2}$ and $U$ the degeneracy is lifted, and $\Gamma(L)$ develops a rather deep minimum for small $U/M^{4}$. The large negative value of $\Gamma_{0}$ is due to the large volume of the sphere. For $U/M^{4}\ll 1$ the geometry of space is almost flat, as observed for the almost de Sitter geometry of our present Universe for Minkoswsi signature. 

In the limit $U/M^{4}\rightarrow 0$ the geometry approaches flat space. In this limit the graviton fluctuations around the configuration~\eqref{CC1},~\eqref{CC5} remain stable. For an expansion around a sphere the graviton fluctuations still correspond to massless excitations with a positive kinetic term. This differs from graviton fluctuations around flat space which develop a tachyonic instability for $U>0$. Besides the graviton, the only physical fluctuation that remains is the scalar fluctuation of the metric. For flat space the effective action contains a term linear in the scalar fluctuation, indicating a different minimum with $\Gamma_{0}<0$. The configuration~\eqref{CC1},~\eqref{CC5} corresponds to this required new extremum for the scalar fluctuation. 

In view of these arguments it seems rather likely that the sphere could be on acceptable minimum for the effective action~\eqref{CC8} for metric gravity within the space of configurations that can be reached by solutions of field equations. For a pure metric theory the propagators derived from this polynomial effective action show tachyonic or ghost instabilities for $q^{2}$ around $M^{2}$. This is of no worry in our case since we have already found stability in this momentum region for the full effective action~\eqref{EA1}.

Relaxing the restriction on configurations with vanishing covariant derivative of the vierbein, it seems useful to consider $e_{\mu}{}^{m}$ and $U_{\mu\nu}{}^{m}$ as variables, instead of $e_{\mu}{}^{m}$  and $A_{\mu mn}$. The term $L_{U}$ provides for a positive quadratic term for $U_{\mu\nu}{}^{m}$. For $n^{2}>-3m^{2}$ this extends to $L_{G}$ . Expanding around the sphere there is no term linear in $U_{\mu\nu}{}^{m}$. The only terms linear in $U_{\mu\nu}{}^{m}$ are from $L_{F}$ via eq.~\eqref{TDI Eq: 40 in F}, and similarly from $L_{G}$. They vanish for a covariantly constant curvature tensor. In principle, mixed terms linear in $U_{\mu\nu}{}^{m}$ and linear in fluctuations of the vierbein could turn the extremum characterizing the sphere to a saddle point. This seems not very likely in the low momentum range, given that the vierbein fluctuations have to generate covariant derivatives of the curvature tensor.

\subsection{Realistic euclidean quantum gravity\label{subsec: Realistic euclidean quantum gravity}}
\smallskip

All these arguments do not constitute a proof that the sphere~\eqref{CC1},~\eqref{CC5} is the minimum of the euclidean effective action~\eqref{EA1} on the space of possible solutions of the field equations. If this sphere is the minimum for the possible solutions of field equations, our formulation of pregeometry leads to a rather satisfactory state of euclidean quantum gravity. The minimum corresponds for tiny $U/M^{4}$ almost to flat space. Its analytic continuation to Minkowski signature would correspond to de Sitter space with a tiny cosmological constant, close to the present Universe for $U/M^{4} \approx 10^{-120}$.

With $U/M^{4}=10^{-120}$, the overall view of the ground state of euclidean pregeometry is a very deep minimum with $\Gamma_{0}\approx -10^{120}$, assumed by a sphere with huge radius in units of the Planck mass, $L_{0}M\approx 10^{60}$. The four dimensional volume of the sphere is extremely large, $L_{0}^{4}M^{4}\approx 10^{240}$, corresponding to the size of our observable Universe in Planck units, $M^{4}/H^{4}\approx 10^{240}$.  Field configurations with even larger volume $L^{4}$ will lead to a huge positive action,  due to the positive term $\int_{x}eU\approx(L/L_{0})^{4}(M^{4}/U)\approx 10^{120}(L/L_{0})^{4}$, which overwhelms all other contributions to $\Gamma$. Fluctuations with effective volume $L^{4}$ substantially smaller than $L_{0}^{4}$ change the effective action only by a small relative amount $\Delta\Gamma/\Gamma_{0}\,$, suppressed by a volume factor $L^{4}/L_{0}^{4}\,$.

\section{Conclusions and discussion\label{sec: Conclusions and discussion}}

In this paper we propose an euclidean functional integral for a model of pregeometry that contains quantum gravity. The classical action is bounded from below and all two-point correlation functions or propagators have a simple short distance behavior without any instabilities. The inverse propagators in flat space are for all physical excitations proportional to the squared momentum $q^{2}$ in the limit $q^{2}\rightarrow\infty$. Analytic continuation to Minkowski signature does not encounter any problems, and there are neither ghost nor tachyon instabilities in the high momentum region. 

The functional integral still needs a regularisation. Since our model of pregeometry is a Yang-Mills gauge theory, the situation is in many respects similar to other gauge theories. The particularity is the diffeomorphism invariance of the classical action that should be preserved by the regularization. Dimensional regularization may be one of the possibilities, needing as all other continuum regularizations the introduction of some type of gauge fixing. It will be interesting to find out if a suitable lattice regularization exists, with $A_{\mu}$ and $e_{\mu}$ replaced by variables on the links of a lattice. This may require lattice diffeomorphism invariance~\cite{CWLDS,CWSG1} for the discrete action, which should extend to the usual diffeomorphism symmetry in the continuum limit. 

We have discussed at length the properties of the quantum effective action and the associated issues of stability of solutions of the field equations. For low momenta it seems reasonable to expect that the invariants consistent with the symmetries and involving a low number of derivatives are present in the quantum effective action. For a realistic theory it is necessary that the term linear in the derivatives of the gauge field is generated by the fluctuations, with a coefficient $M^{2}>0$. Furthermore, the couplings in a derivative approximation to the effective action have to be in a range for which all fluctuations around flat space are stable for a vanishing covariant derivative of $e_{\mu}\ $,  $U_{\mu\nu\rho}=0$. In this case general relativity with the Einstein-Hilbert action for the composite metric emerges as the effective low momentum theory. By analytic continuation our model of pregeometry can then describe quantum gravity. 

It is straightforward to add fermions in our model. The corresponding Grassmann variables $\psi$ are scalars with respect to general coordinate transformations and transform as Weyl or Dirac spinors with respect to the $SO(4)$ - gauge group. The covariant derivative contains the gauge field in the usual way. The gauge invariant kinetic term for the fermions reads 
\begin{align}\label{CD01}
S_{\psi}=i\int_{x} e\overline{\psi}\gamma^{m}e_{m}{}^{\mu}D_{\mu}\psi\;,\quad \overline{\psi}=\psi^{\dagger}\gamma^{0}\,,\nn\\
D_{\mu}\psi=\partial_{\mu}\psi-\dfrac{1}{2}A_{\mu m n}\Sigma^{mn}\psi\;,
\end{align}
with $\gamma^{m}$ the euclidean Dirac matrices and $\Sigma^{mn}$ the $SO(4)$ - generators in the spinor representation, 
\be\label{CD02}
\Sigma^{mn}=-\dfrac{1}{4}\bigl[\gamma^{m},\gamma^{n}\bigr]\;,\quad \bigl{\lbrace}\gamma^{m},\gamma^{n}\bigr{\rbrace}=2\delta^{mn}\;.
\ee
With $ee_{m}{}^{\mu}$ being a polynomial in the vierbein, 
\be\label{CD03}
ee_{m}{}^{\mu}\sim\varepsilon^{\mu\nu\rho\sigma}\varepsilon_{mnpq}\, e_{\nu}{}^{n}e_{\rho}{}^{p}e_{\sigma}{}^{q}\;,
\ee
this action does actually not involve the inverse vierbein. Weyl spinors obtain by suitable projections, $\psi_{\pm}=\big{(}(1\pm \gamma^{5})/2\big{)}\psi\ $, $\gamma^{5}=\eta\ \!\gamma^{0}\gamma^{1}\gamma^{2}\gamma^{3}$.

It is possible that the vierbein arises from an even more fundamental pregeometry as a composite fermion bilinear
\be\label{CD04}
e_{\mu}{}^{m}=i\overline{\psi}\gamma^{m}D_{\mu}\psi\;.
\ee
Insertion of this expression into eqs.~\eqref{CD01},~\eqref{CD03} yields an invariant term involving eight powers of the fermion fields $\psi$. This is a gauge theory based only on fermions and gauge fields. Finally, the gauge fields may also be expressed as composites of fermions similar to ref.~\cite{CWSLG}.
This results in a type of spinor gravity with local ``Lorentz" symmetry $SO(4)$ and diffeomorphism invariance~\cite{CWSG1,CWSG2}. While conceptually rather interesting, it is clear that suitable technologies have to be developed for dealing with actions that involve only multi-fermion invariants.

In the direction of a more realistic model for particle physics coupled to gravity, one needs to include additional gauge symmetries and corresponding gauge fields, as for the standard model gauge symmetry $SU(3)\times SU(2)\times U(1)$ or some grand unified extension. Furthermore, scalar fields are needed for a description of spontaneous symmetry breaking. Interesting cosmologies can be obtained if the parameters $Z$, $\tilde{Z} $,  $m^{2}$, $\tilde{m}^{2}$, $M^{2}$ become functions of a scalar singlet field $\chi$, similar to variable gravity~\cite{CWVG}. A first investigation~\cite{CWPC} reveals that rather realistic cosmology with inflation and dynamical dark energy can be obtained for a suitable  shape of the coupling functions.

The central remaining issue with which we have not addressed in this paper concerns the renormalizability of our model of pregeometry. This involves an investigation of the dependence of the couplings $Z$, $C$, $m^{2}$,$n^{2}$,  $M^{2}$ and $U$ on some renormalization scale $k$. More generally, beyond the couplings mentioned explicitly, one needs to understand the dependence of the whole scale-dependent effective action $\Gamma_{k}$ on $k$. This question is a typical issue for functional renormalization which can be addressed by suitable truncations of the exact flow equation for the effective average action~\cite{CWFE}. The scale-dependent effective action should admit a fixed point or scaling solution for $k\rightarrow\infty$, which remains sufficiently simple and close to the proposed classical action.
Only in this case our model of pregeometry can be considered as a fully satisfactory description of emerging quantum gravity.
\medskip

\bibliography{refs}

\end{document}